\documentclass[%
twocolumn,
nofootinbib,
 unsortedaddress,
 amsmath,amssymb,                
]{revtex4-2}

\usepackage{graphicx}
\usepackage{dcolumn}
\usepackage{booktabs,siunitx}
\usepackage{xcolor}
\usepackage{physics}
\usepackage{bbold}
\usepackage{nicefrac}
\usepackage{bm}

\renewcommand{\onlinecite}[1]{\hspace{-1 ex} \nocite{#1}\citenum{#1}}

\begin{document}
\setcitestyle{super}

\title{Temperature- and vacancy-{concentration-}dependence of heat transport in Li\textsubscript{3}ClO from 
multi-method numerical simulations}

\author{Paolo Pegolo}
\affiliation{SISSA -- Scuola Internazionale Superiore di Studi Avanzati, 34136 Trieste, Italy.}
\author{Stefano Baroni}
\affiliation{SISSA -- Scuola Internazionale Superiore di Studi Avanzati, 34136 Trieste, Italy.}
\author{Federico Grasselli}\email{federico.grasselli@epfl.ch}
\affiliation{COSMO -- Laboratory of Computational Science and Modelling, IMX, \'Ecole Polytechnique F\'ed\'erale de Lausanne, 1015 Lausanne, Switzerland,}
\affiliation{SISSA -- Scuola Internazionale Superiore di Studi Avanzati, 34136 Trieste, Italy.}

\begin{abstract}
Despite governing heat management in any realistic device, the microscopic mechanisms of heat transport in all-solid-state electrolytes are poorly known: existing calculations, all based on simplistic semi-empirical models,  are unreliable for superionic conductors and largely overestimate their thermal conductivity. In this work, we deploy a combination of state-of-the-art methods to calculate the thermal conductivity of a prototypical Li-ion conductor, the Li\textsubscript{3}ClO antiperovskite. By leveraging \emph{ab initio}, machine-learning, and force-field descriptions of inter-atomic forces, we are able to reveal the massive role of anharmonic interactions and diffusive defects on the thermal conductivity and its temperature dependence, and to eventually embed their effects into a simple rationale which is likely applicable to a wide class of ionic conductors.
\end{abstract}

\maketitle

\section{Introduction}

The lithium-rich antiperovskite solid-state electrolyte (SSE) Li\textsubscript{3}ClO has emerged in the last decade as a promising candidate for all-solid-state lithium-metal batteries: it is superionic at room temperature with a large ionic conductivity; it is environmental friendly and made of light and cheap elements; it is not flammable and has demonstrated a good cyclability.~\cite{braga2014novel} Furthermore, its wide electronic band gap leads to a very low electronic conductivity and a large electro-chemical stability window.~\cite{lu2014li} Finally, it is also chemically stable against Li-metal formation, which would negatively affect the battery performance via dendritic short-circuits.

A proper account of heat dissipation is key in the design and actual production of batteries: in fact, an excessively low thermal conductivity may lead to overheating, especially during fast charging cycles, which may itself prompt catastrophic incidents, such as melting or explosion.  In light of this, thermal runaway can be rightly considered ``the key scientific problem in battery safety research'' (\emph{verbatim} from Ref.~\onlinecite{feng2018thermal}). In addition, thermal dissipation governs energy saving and scavenging: a compromise must be established between minimising heat losses while maximising electric flow during the charging cycle.

A relatively small number of experimental studies on heat dissipation in fast ionic conductors have been reported, as compared to the extensive literature on electric transport. Measurement were performed on superionic materials such as $\alpha$-$\mathrm{LiIO_3}$ and $\mathrm{Li_2B_4O_7}$;~\cite{Abdulchalikova1995thermal, aliev1997specific} sulfates $\mathrm{Li_2SO_4}$ and $\mathrm{Ag_2SO_4}$;~\cite{elrahman1999electrical} the quasi-1D material $\mathrm{LiCuVO_4}$;~\cite{parfen2003heat} lithium aluminium germanium phosphate glass–ceramics compounds of the form $\mathrm{Li}_{1+x}\mathrm{Al}_x\mathrm{Ge}_{2-x}\mathrm{(PO_4)_3}$;~\cite{cui2016thermal} yttrium-stabilised lithium zirconate phosphates, of the form $\mathrm{Li}_{1+x+y}\mathrm{Y}_x\mathrm{Zr}_{2-x}\mathrm{(PO_4)_3}$.~\cite{yazdani2020thermal} Of these, only the last two classes of materials can can be directly used as SSEs. From the theoretical standpoint, even fewer works are to be found. Even though some models that account for the contribution of the diffusing ions to thermal transport were introduced in the past decades,~\cite{rice1972ionic, yonashiro1988thermal} a thorough study of heat dissipation in Li\textsubscript{3}ClO---and in SSE in general---is still missing. At the best of our knowledge, there exists only one calculation of the thermal conductivity, $\kappa$, of Li\textsubscript{3}ClO, reporting ${\kappa = 22.49\,\mathrm{W\,m^{-1}\,K^{-1}}}$ at ambient temperature, \emph{i.e.}, more than one order of magnitude larger than the standard value found in ceramic SSEs.~\cite{cui2016thermal, yazdani2020thermal} Nonetheless, this seemingly promising result is obtained via a rather crude approximation to the Peierls-Boltzmann transport equation (BTE), namely the Slack model, known since its development to have a satisfactory agreement with the experiments (\emph{i.e.},~within $\pm 20$\%) only for exceedingly simple materials, such as the rare-gas solids, while it is in general poorer for other systems.~\cite{slack1979thermal} Furthermore, the Slack approximation totally neglects the effects of defects/vacancies and, just like any BTE-based model, it cannot handle the spurious contributions to heat transport induced by the diffusion of Li ions. These limitations, stemming from a phonon(/normal-mode)-based approach to heat transport, can be naturally and easily bypassed through the Green-Kubo (GK) theory of linear response,~\cite{green1952markoff,green1954markoff,kubo1957statistical1,kubo1957statistical2,marcolongo2016microscopic,isaeva2019modeling} which holds for solids with anharmonic interactions of any strength, as well as for diffusing systems, such as liquids and superionic solids. Recently, the GK theory of thermal transport has been combined with state-of-the art quantum simulation methods based on density-functional theory (DFT) ~\cite{marcolongo2016microscopic,baroni2020heat,grasselli2021invariance} and successfully employed to evaluate $\kappa$ from equilibrium molecular dynamics simulations of superionic systems, even at extreme pressure and temperature conditions.~\cite{grasselli2020heat, stixrude2021thermal}

In this paper we first address the structural and mechanical properties of Li\textsubscript{3}ClO, along with their temperature dependence, by means of state-of-the-art \emph{ab initio} calculations based on DFT. Leveraging these results, we then explore the thermal transport of Li\textsubscript{3}ClO by calculating its thermal conductivity within \textit{i)} the Slack model; \textit{ii)} a state-of-the art implementation of the BTE; and \textit{iii)} the GK theory of linear response and classical molecular dynamics (MD) simulations using both classical force fields and machine-learnt interatomic potentials, trained on DFT data. Our results show that the presence of {LiCl} divacancies {in non-stoichiometric systems}, though increasing the Li-ion diffusivity, strongly reduces the thermal conductivity---and thus heat dissipation within the electrolyte---with respect to {stoichiometric} conditions. We find that the dependence of $\kappa$ on temperature is also reduced, thus making it potentially easier to engineer devices that can safely and efficiently operate in a wide range of temperatures.

\section{Results}\label{sec:results}

Before addressing thermal transport (Sec.~\ref{sec:TC}), which represents the main goal of our work, we investigate the structural (Sec.~\ref{sec:SP}), electronic (Sec.~\ref{sec:EBS}), vibrational (Sec.~\ref{sec:PH}), and mechanical (Sec.~\ref{sec:MP}) properties of Li\textsubscript{3}ClO, along with their temperature dependence, and we extensively compare our \emph{ab initio} results with the existing literature. These results are not only preliminary to the calculation of transport coefficients, but they also make it possible to draw some general conclusions on the deployment of Li\textsubscript{3}ClO for mass production of SSE-based batteries: ductility, stiffness, the magnitude of the mismatch in the lattice constants or in the thermal expansion coefficients between the SSE and the electrodes govern the extent and amplitude of local stresses, particularly at the SSE-electrode junction, and affect the overall performance of Li\textsubscript{3}ClO in a real device by either hindering or inducing cracks, mechanical instabilities, or spurious electric fields.

\subsection{Structural properties}\label{sec:SP}
\begin{figure}[htb]
    \centering
    \includegraphics[width = \columnwidth]{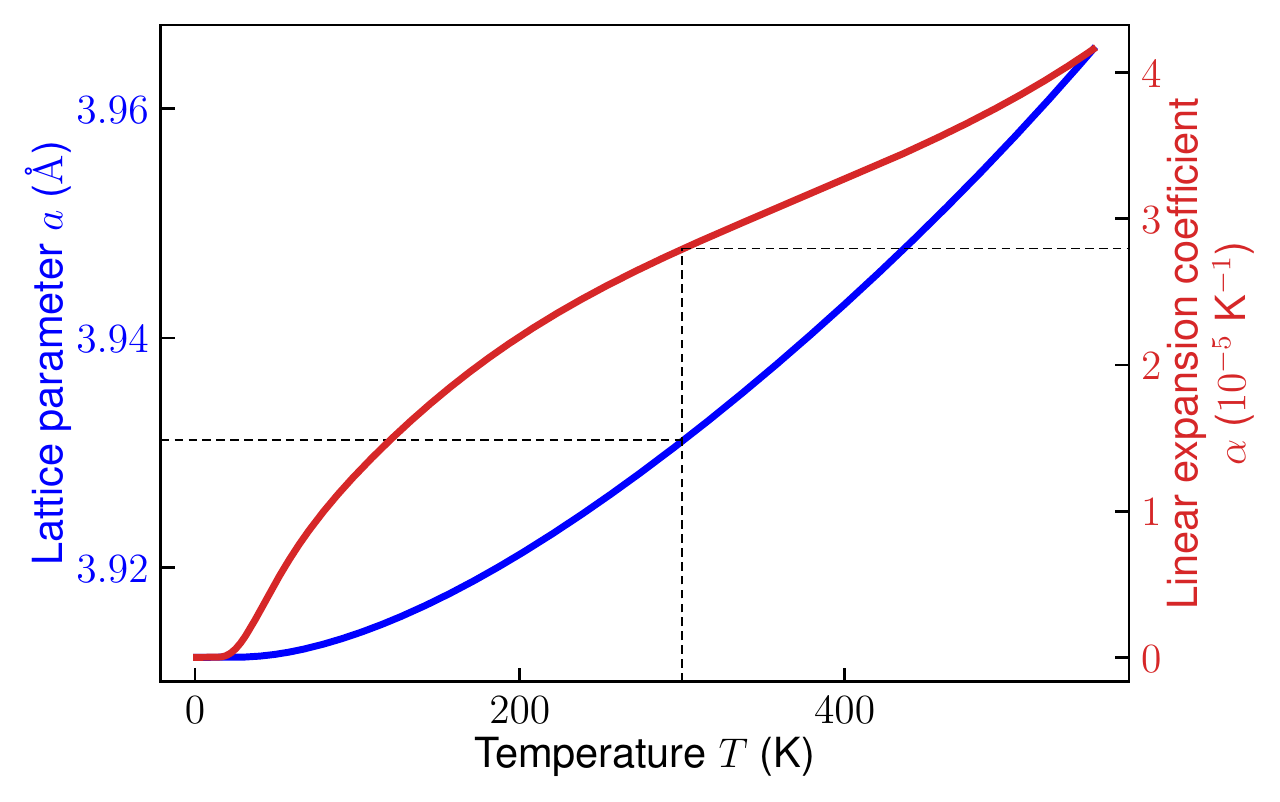}
    \caption{QHA structural parameters. Lattice parameter (blue, left $y$-label) and linear thermal expansion coefficient (red, right $y$-label) as a function of temperature.}
    \label{fig: alat and alpha vs t}
\end{figure}

The crystal structure of Li\textsubscript{3}ClO has the  $\mathrm{Pm\overline{3}m}$ perovskite space group. The dependence of the lattice parameter on temperature is obtained in the quasi-harmonic approximation (QHA),~\cite{Baroni2010QHA} using the the vibrational frequencies computed as explained in Sec.~\ref{sec:PH} and the Murnaghan equation of state.~\cite{Murnaghan244} The resulting zero-temperature value, explicitly accounting for zero-point vibrational effects, is $3.91\,$\AA, while the full temperature dependence is reported in Fig.~\ref{fig: alat and alpha vs t}.
The lattice parameter and thermal expansion coefficient computed at room temperature ($\mathrm{RT}=300\,\mathrm{K}$) are $3.93\,\text{\AA}$ and $2.77 \cdot 10^{-5}\,\mathrm{K^{-1}}$, respectively, in good agreement with the values previously obtained in other works: Zhang \emph{et al.},~\cite{Zhang2013} Emly \emph{et al.},~\cite{emly2013phase} and Wu \emph{et al.}~\cite{wu2018bulk} reported an optimised lattice parameter without the zero-point of ${3.85}\,$\AA, $3.90\,$\AA, and ${3.91}\,$\AA, respectively; Braga \emph{et al.}~\cite{braga2014novel} provided both experimental and theoretical data in accordance with one another, namely $3.91\,$\AA.~\cite{zhao2012superionic, braga2014novel} For the linear thermal expansion coefficient, Zhang \emph{et al.}~\cite{Zhang2013} reported a value of $2.11\times10^{-5}\,\mathrm{K^{-1}}$ using molecular dynamics simulations in the NPT ensemble, while Wu \emph{et al.}~\cite{wu2018bulk} reported  $\alpha=3.12\times10^{-5}\,\mathrm{K^{-1}}$ within the quasi harmonic approximation (QHA), in good agreement with our calculation. Braga \emph{et al.}~\cite{braga2014novel} found the larger value $4.65\times10^{-5}\,\mathrm{K^{-1}}$ from the slope of the lattice parameter as a function of temperature.

\subsection{Electronic band structure}\label{sec:EBS}

\begin{figure}[htb]
    \centering
    \includegraphics[width=\columnwidth]{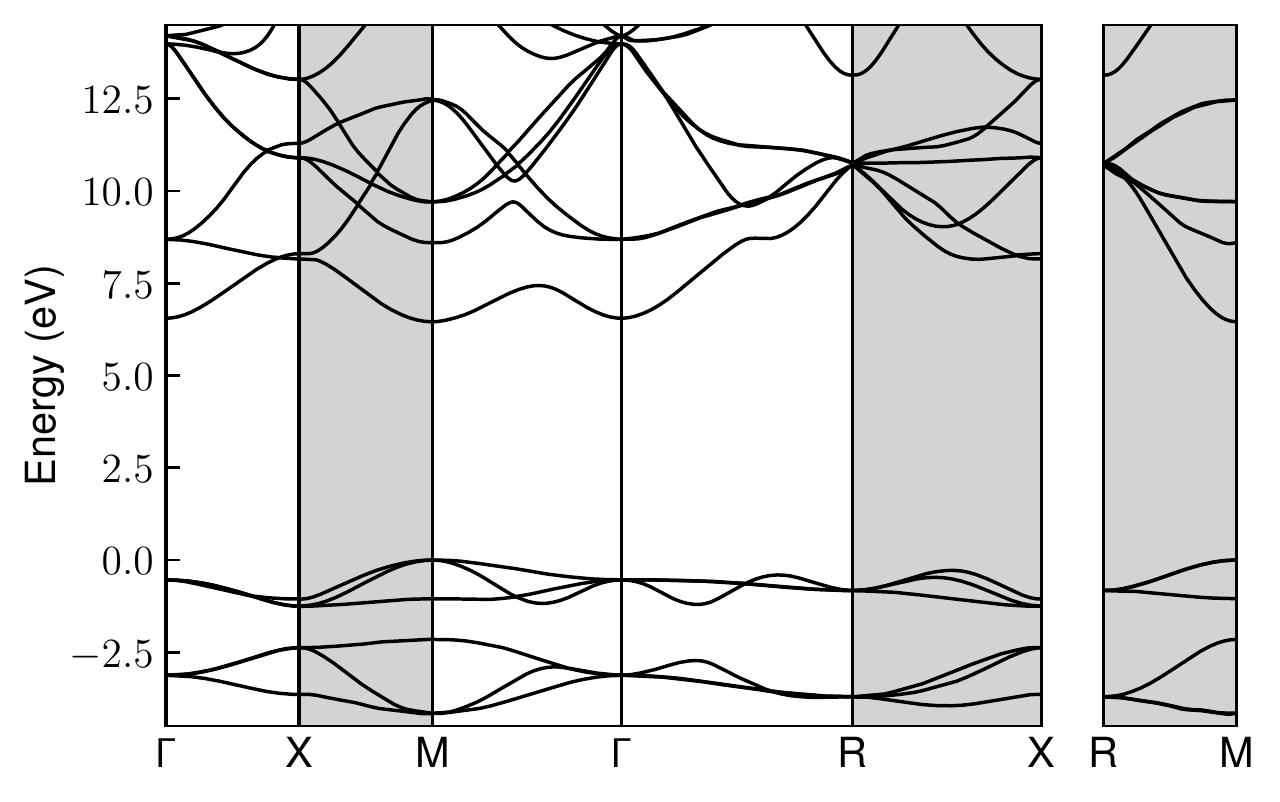}
    \caption{Electronic band dispersion of Li\textsubscript{3}ClO plotted along high-symmetry lines in the BZ. Grey-shaded areas represent lines that lay on the BZ border.}
    \label{fig: bands}
\end{figure}

The electronic band structure of Li\textsubscript{3}ClO (Fig.~\ref{fig: bands}) exhibits a direct band gap of $6.46\,\mathrm{eV}$ at the M in the BZ at the HSE06 level of theory, in good agreement with the values obtained by Wu \textit{et al.}~\cite{wu2018bulk} ($6.46\,\mathrm{eV}$) and by Emly \textit{et at.}~\cite{emly2013phase}, thus making Li\textsubscript{3}ClO a wide band-gap insulator and preluding a good electrochemical stability.

\subsection{Vibrational properties}\label{sec:PH}

\begin{figure}[htb]
    \centering
    \includegraphics[width=\columnwidth]{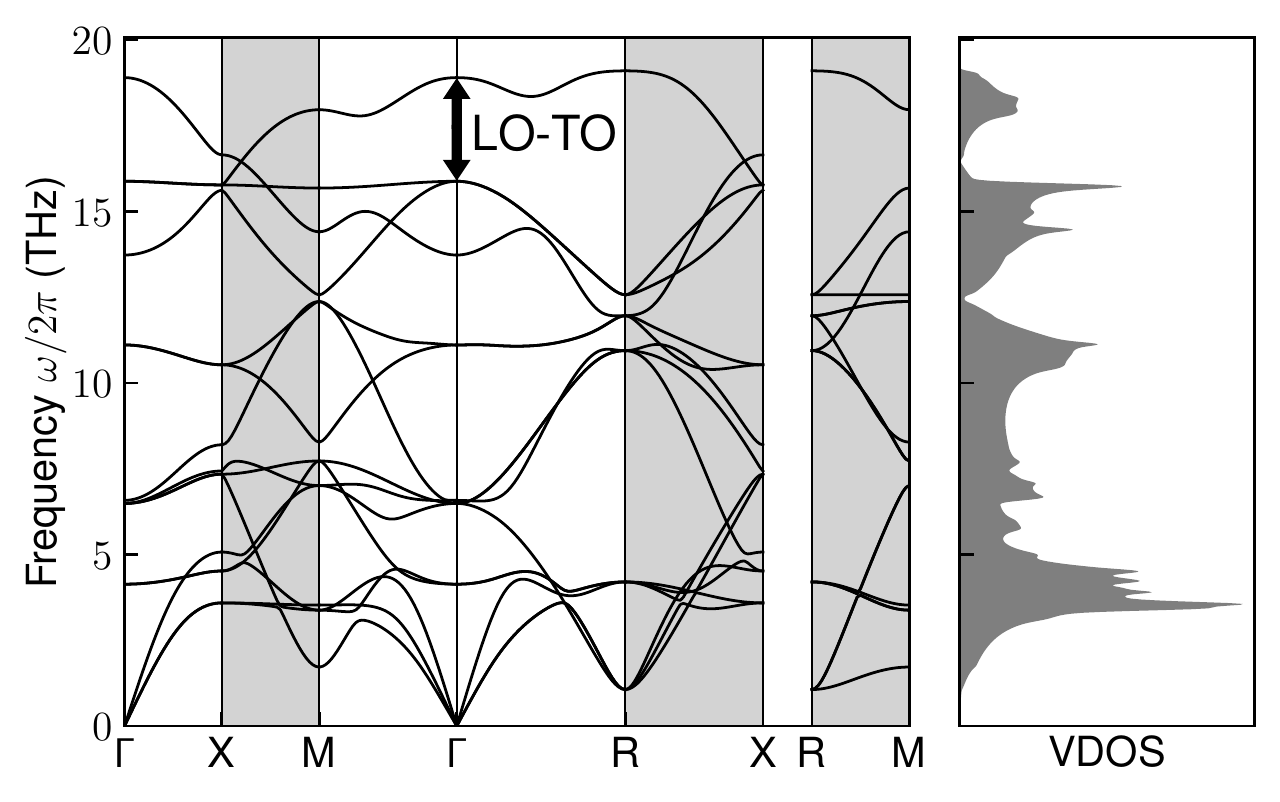}
    \caption{Phonon dispersion of Li\textsubscript{3}ClO plotted along high-symmetry lines in the BZ and respective density of states. Grey-shaded areas represent lines that lay on the BZ border. The LO-TO splitting at $\Gamma$ is evident.
    }
    \label{fig: phonons}
\end{figure}

The phonon dispersion along high-symmetry lines in the BZ are plotted in Fig.~\ref{fig: phonons} together with the corresponding vibrational density of states (VDOS). Notice the presence of a large longitudinal-transverse (LO-TO) splitting in the infrared-active mode at the centre of the BZ ($\Gamma$ point), due to the non-analytic behaviour of the dynamical matrices induced by the long-range (Coulomb) tails of the inter-atomic interactions. It has been pointed out that a proper account of these tails is  essential not only for a correct qualitative description of the vibrational spectrum in the optic region, but also for an accurate evaluation of the thermal conductivity, which is more sensitive to the low-frequency portion of the spectrum.~\cite{shafique2020effect} The effect of a proper account of the LO-TO splitting on the value of the lattice thermal conductivity of Li\textsubscript{3}ClO is examined in Appendix~\ref{ssec:NAC}.

The stability at zero temperature of this material has been debated, with different authors claiming the system to be either unstable~\cite{chen2015anharmonicity} or stable~\cite{wu2018bulk}: methods are employed, the authors find soft modes (imaginary frequencies) at the $\mathrm{M}$ and $\mathrm{R}$ points in the BZ using a ${6 \times 6 \times 6}$ supercell. Since the instability is larger when a smaller (${3 \times 3 \times 3}$) supercell is used, the occurrence of lattice instabilities may be an artefact due to lack of convergence even when a supercell as large as ${6 \times 6 \times 6}$ unit cells is used in finite-difference calculation. Using perturbation theory at the zone-center of a supercell,\footnote{This by construction does not include NAC.} the authors of Ref.~\onlinecite{wu2018bulk} do not find any dynamical instabilities.
Our calculations, performed within density functional perturbation theory (DFPT), confirm the dynamical stability of this system.

\subsection{Temperature dependence of the mechanical properties}\label{sec:MP}

Mechanical properties can have a deep influence on the fabrication of any device and, in particular, of alkali-ion solid-state batteries. An important aspect is that there must be a good contact between electrolyte and electrodes during the activity of a device. A good SSE candidate must be able to sustain large strains~\cite{deng2015elastic} to prevent the interfaces with cathode and anode to deteriorate in response to the deformation thereof. In this light, excessive stiffness is a feature to be avoided. Another element to consider is the problem of Li deposition at the interface with the cathode, and the subsequent dendrite formation,~\cite{XuLithium2014} that affects especially liquids. A solid material may partially overcome this obstacle by fine-tuning elastic properties such as the shear modulus and Poisson's ratio,~\cite{monroe2005impact} even if it has been reported that dendrite growth can still occur for other reasons.~\cite{dolle2002live}

\begin{figure*}[tb]
    \centering
    \includegraphics[width=2\columnwidth]{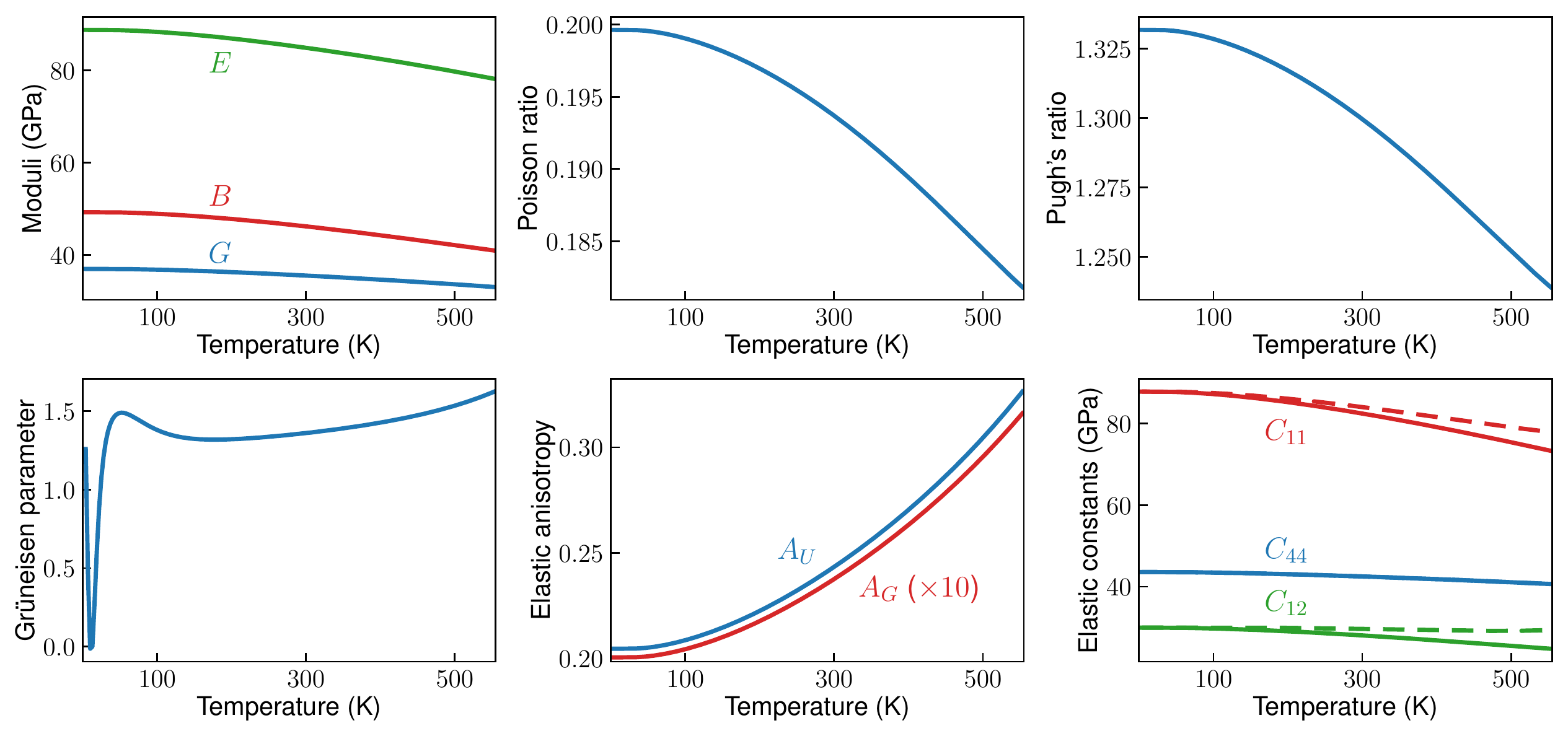}
    \caption{Temperature dependence of the macroscopic mechanical properties of Li\textsubscript{3}ClO. The elastic constants are computed at the QSA level; solid lines are isothermal elastic constants, while dashed lines are isoentropic. All the other quantities are computed according to the QHA.}
    \label{fig: macroscopic mechanical}
\end{figure*}

The three elastic moduli, namely the bulk modulus $B$, the Young modulus $E$, and the shear modulus $G$, measure how a material responds to volumetric, tensile, and shear stress, respectively; Poisson's ratio, $\nu$, measures the strain response in a direction perpendicular to an applied strain; Pugh's ratio, $B/G$, is related to how materials are ductile as opposed to brittle. The Chung-Buessem~\cite{chung1967elastic} elastic anisotropy index, $A_G$, and the universal elastic anisotropy,~\cite{ranganathan2008universal} $A_U$, provide a measure to quantify the extent of anisotropy in the elastic response of a crystal, and are obtained from the Voigt and Reuss estimates of $B$ and $G$.~\cite{ranganathan2008universal} The Gr\"uneisen parameter $\gamma$ is the variation of pressure with thermal energy density at constant volume.~\cite{stacey2019thermodynamics} All these quantities are computed as a function of temperature within the QHA. The isothermal and adiabatic elastic constants ($C_{ij}$) as a function of temperature are computed within the quasi-static approximation~(QSA).

The temperature dependence of these quantities is shown in Fig.~\ref{fig: macroscopic mechanical}. Their values at RT and at $T=0$ are reported in Table~\ref{tbl: mechanical properties}, together with results already available in the literature.~\cite{deng2015elastic, Zhang2013} While our results at $T=0$ are comparable with the ones found in other works, at the working temperature the quantities have lower values, in accordance to the fact that the material becomes softer when temperature is increased. In particular, the temperature dependence of $B$ and $G$ entails that Pugh's ratio is also decreased, hinting at a greater brittleness~\cite{pugh1954xcii} of Li\textsubscript{3}ClO than previously predicted.~\cite{deng2015elastic, wu2018bulk} Furthermore, the computed value of $A_G$ is higher than what found in the literature, suggesting that Li\textsubscript{3}ClO is slightly more anisotropic than previously thought; this is confirmed by the value of $A_U$, that allows a broader comparison with other materials.

\begin{table}[h]
\small
  \caption{Macroscopic mechanical properties of Li\textsubscript{3}ClO at RT. For comparison, we report data from other calculations done at zero temperature.
  }
  \label{tbl: mechanical properties}
  \begin{tabular*}{\columnwidth}{@{\extracolsep{\fill}}l *{4}{S[table-format=3.3]}}
    \toprule
    & \multicolumn{1}{c}{This work (RT)} & \multicolumn{1}{l}{This work ($T=0$)} & \multicolumn{1}{l}{Ref.~\onlinecite{deng2015elastic}} & \multicolumn{1}{l}{Ref.~\onlinecite{wu2018bulk}}\\
    \midrule
    $B\,(\mathrm{GPa})$ & 46.01 & 48.87 & 55.70 & 51.36\\
    $E\,(\mathrm{GPa})$ & 84.86 & 88.65 & 99.70 & 91.93 \\
    $G\,(\mathrm{GPa})$ & 35.58 & 37.01 & 41.50 & 38.25\\
    $\nu$ & 0.19 & 0.20 & 0.20 & 0.20 \\
    $B/G$ & 1.29 & 1.32 & 1.35 & 1.35 \\
    $A_G$ & 0.024 & 0.020 & {\textemdash} & 0.01 \\
    $A_U$ & 0.24 & 0.20 & {\textemdash} & {\textemdash} \\
    $\gamma$ & 1.30 & 1.86 & {\textemdash} & {\textemdash} \\
    $C_{11}\,(\mathrm{GPa})$ & 82.32 & 87.47 & 102.90 & 93.68 \\
    $C_{12}\,(\mathrm{GPa})$ & 27.85 & 29.57 & 32.10  & 30.20\\
    $C_{44}\,(\mathrm{GPa})$ & 42.55 & 43.62 & 46.10  & 43.33\\
    \bottomrule
  \end{tabular*}
\end{table}

\subsection{Thermal conductivity}\label{sec:TC}

Thermal conductivity ($\kappa$) plays a key role in the quest for promising SSE for battery production. A high value of the thermal conductivity is a favourable quality for a SSE candidate to have, since efficient heat dissipation allows the manufacture of safer batteries that do not overheat. A quite high value $\kappa = 22.49\,\mathrm{Wm^{-1}K^{-1}}$ was previously reported.~\cite{wu2018bulk} This result was obtained from Slack's model (SM),~\cite{slack1979thermal} which is based on a rather crude approximation of the BTE~\cite{peierls1929kinetischen, ziman2001electrons, mcgaughey2019phonon}. We briefly review SM in Appendix~\ref{app: slack}. At the same level of theory, we obtain $16.55\,\mathrm{W\,m^{-1}\,K^{-1}}$ that is 26\% lower and it is likely to be closer to a realistic value. Yet, the coarseness of the approximation makes it impossible to draw conclusions: as already mentioned in the Introduction, SM is known, since its formulation dating back to 1960s, to be fairly accurate (to within 20\%) only for simple systems such as rare-gas crystals, while it grossly fails for more complex materials. The main issue is to be addressed to the cubic dependence of $\kappa$ on the Debye temperature $\Theta_D$ of the crystal, which itself depends dramatically on the parameters used to estimate it. For instance, if the lattice parameter is reduced by less than 2\%, \emph{i.e.}, from the value we compute ($3.91\,\text{\AA}$) to a value found in the literature~\cite{Zhang2013} ($3.85\,\text{\AA}$), $\Theta_D$ increases by 4\% (from $630\,\mathrm{K}$ to $653\,\mathrm{K}$) and $\kappa$ by 12\% (from $16.6\,\mathrm{W\,m^{-1}\,K^{-1}}$ to $18.5\,\mathrm{W\,m^{-1}\,K^{-1}}$). This and further approximations of the model---it is for instance insensible to the crystal structure of the material---makes it impossible to produce reliable estimates for $\kappa$ within SM. For a realistic calculation the BTE must be treated in a detailed fashion relying on the explicit calculation (and inversion) of the phononic scattering-rate matrix to obtain the out-of-equilibrium occupation numbers of the phonons. A quick review of the BTE-based approach we follow is given in Appendix~\ref{app: BTE}.

\begin{figure}
    \centering
    \includegraphics[width=\columnwidth]{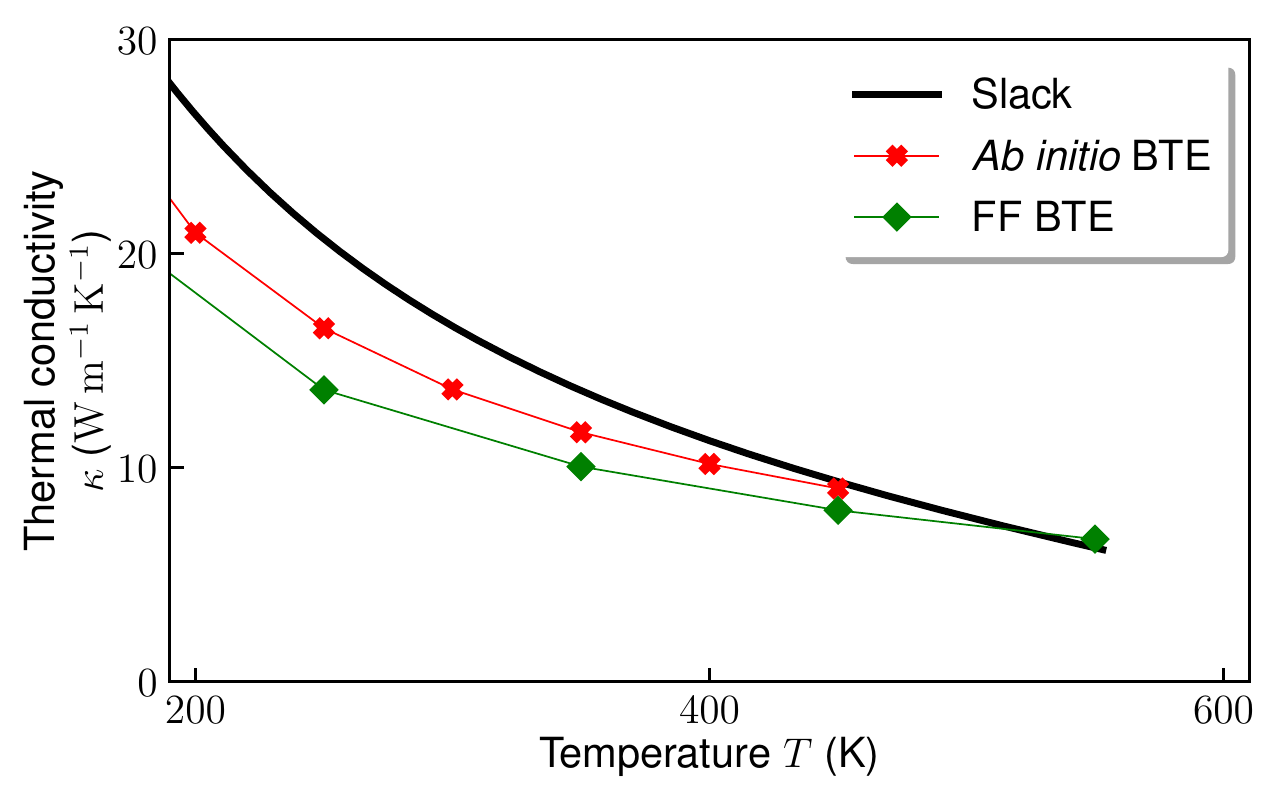}
    \caption{Comparison of lattice thermal conductivities given by different methods. BTE calculations are within SMRTA with 3-phonon scattering. Unfortunately, due to the large anharmonicity, 3-phonon scattering is not sufficient and further terms must be included, as discussed in the main text and shown quantitatively in Fig.~\ref{fig: 4 phonon}.}
    \label{fig: kappa lattice comparison}
\end{figure}

\begin{figure}[htb]
    \centering
    \includegraphics[width=\columnwidth]{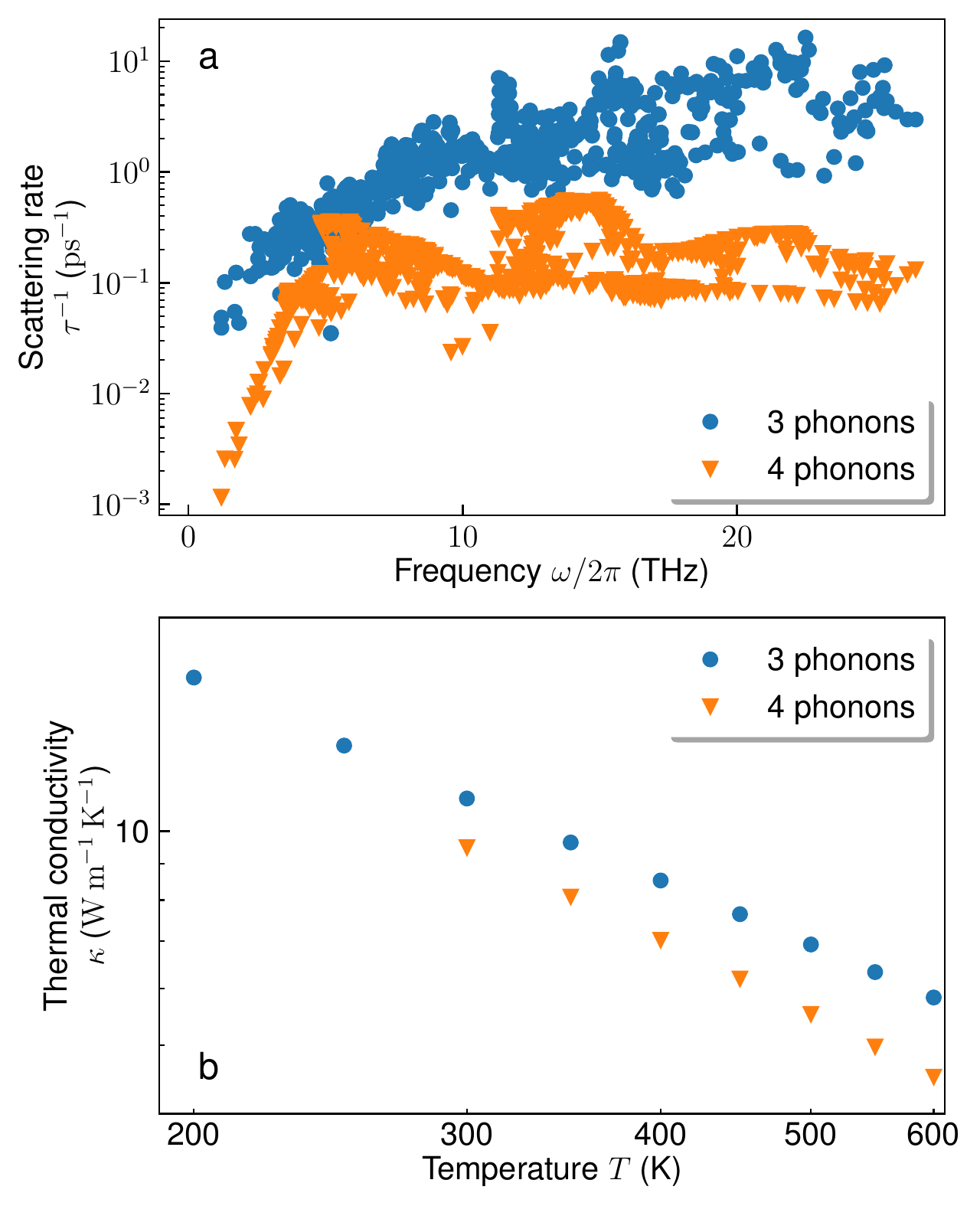}
    \caption{Effect of the inclusion of four-phonon scattering. (\emph{up}) The four-phonons scattering rates---shown here are at RT---are non-negligible with respect to the three-phonon contribution. (\emph{down}) Lattice thermal conductivity is consequently reduced.}
    \label{fig: 4 phonon}
\end{figure}

The lattice thermal conductivity in the Single Mode Relaxation Time Approximation (SMRTA) is computed \emph{ab initio} via genuine third-order perturbation theory as implemented in the \texttt{D3Q} code~\cite{paulatto2013anharmonic} distributed with \textsc{Quantum~ESPRESSO}, and is displayed in Fig.~\ref{fig: kappa lattice comparison} as a function of the temperature. For a comparison, we also report our calculation within the SM, which agrees qualitatively to the \emph{ab initio} SMRTA result better than the existing literature.~\cite{wu2018bulk} We test the reliability of the classical FF in probing thermal properties of Li\textsubscript{3}ClO by employing it to compute the lattice thermal conductivity in the SMRTA. In general, a full solution of the linearised BTE may change significantly the computationally cheaper SMRTA result. In practice, we explicitly verified that the full solution reduces the SMRTA value of up to $0.5\%$, so we can safely keep all the calculations at the SMRTA level. Details can be found in Appendix~\ref{ssec:full-bte-vs-rta}. Figure \ref{fig: kappa lattice comparison} shows that classical FF and AI results are in good agreement, suggesting that the chosen classical FF is suited for investigating the thermal transport properties of Li\textsubscript{3}ClO.
As recently pointed out,~\cite{feng2016quantum, feng2017four} higher-order scattering events can drastically reduce the value of the lattice thermal conductivity; therefore, we test the effect of the inclusion of four-phonon scattering into the computation of $\kappa$. As shown in Fig.~\ref{fig: 4 phonon}, four-phonon scattering is found to have a major role in determining the value of the lattice thermal conductivity, being able to reduce $\kappa$ of at least $15\%$ at RT, the reduction being larger for larger temperatures. Details on the calculation can be found in Appendix~\ref{ssec:4-phonons}. This facts calls for a method able to include higher-order scattering: molecular dynamics (MD) simulations together with GK linear response theory (see Sec.~\ref{sec: GK}), which automatically include all the orders of interactions among phonons, are a fitting candidate for this role. \footnote{
Notice that, in BTE, the occupation numbers follow the Bose-Einstein distribution, $n=\left(\mathrm{e}^{\nicefrac{\hbar\omega}{k_\mathrm{B}T}}-1\right)^{-1}$,  while, in MD, $n=\nicefrac{k_\mathrm{B}T}{\hbar\omega}$, according to classical equipartition. This always leads to an underestimation of thermal conductivity in MD with respect to BTE, at the same level of the theory. An account on the relationship between GK and BTE approaches to thermal transport can be found in Ref.~\onlinecite{puligheddu2019computational}. Further details can be found in Appendix~\ref{ssec:equipartition-vs-be}.}

The choice of MD comes with an additional benefit: since fast ion conduction in SSE materials such as Li\textsubscript{3}ClO is due to diffusing defects, lattice dynamical methods are inadequate, as they require the atoms to have fixed equilibrium positions. MD simulations are not subject to this prerequisite and, therefore, they allow us to access thermal transport in the diffusive regime where fast ion diffusion is mediated by vacancy hopping.

According to the comprehensive studies of Mouta \emph{et al}~\cite{mouta2014concentration} and Lu \emph{et al},~\cite{lu2015defect} $\mathrm{LiCl   }$ Schottky pairs---divacancies generated by the removal of neutral groups of atoms that are deposited at the surface of the material---are more likely to appear and give rise to a high Li-ion mobility than both other Schottky ($\mathrm{Li_2O}$ or Li\textsubscript{3}ClO vacancies) and Frenkel (Li vacancies and interstitials) defects. Thus, the nonstoichiometric systems we study are of the form $\mathrm{Li}_{3-x}\mathrm{Cl}_{1-x}\mathrm{O}$, with $x$ the concentration of vacancies that varies between $0$ (perfect crystal) and $0.1$.

The classical FF described in Sec.~\ref{sec: classical FF} is used to carry out the MD simulations and sample the heat flux according to Eq.~\eqref{eq: J_E}, which in turn is employed to compute $\kappa$ as in Eq.~\eqref{eq: kappa_multi}. To validate this, we compare classical FF calculations to estimates of $\kappa$ extracted from \emph{i)} a Car-Parrinello MD simulation, on which a DFT energy flux is computed according to Ref.~\onlinecite{marcolongo2016microscopic} as implemented in Ref.~\onlinecite{marcolongo2021qeheat}; \emph{ii)} a model obtained via machine learning techniques.

The \emph{ab initio} GK thermal conductivity requires a computationally demanding DFT energy flux (it requires roughly twice the computational time of the AIMD simulation the trajectory is sampled from): therefore, a single calculation at RT is carried out for the sake of checking the accuracy of the FF. We obtain ${\kappa_{\mathrm{DFT}}=6.5 \pm 1.0\,\mathrm{W\,m^{-1}\,K^{-1}}}$ at ${T=306\,\mathrm{K}}$ on a ${4 \times 4 \times 4}$ supercell of crystalline Li\textsubscript{3}ClO, in close agreement with results from classical FF (see below).

\begin{figure}
    \centering
    \includegraphics[width=\columnwidth]{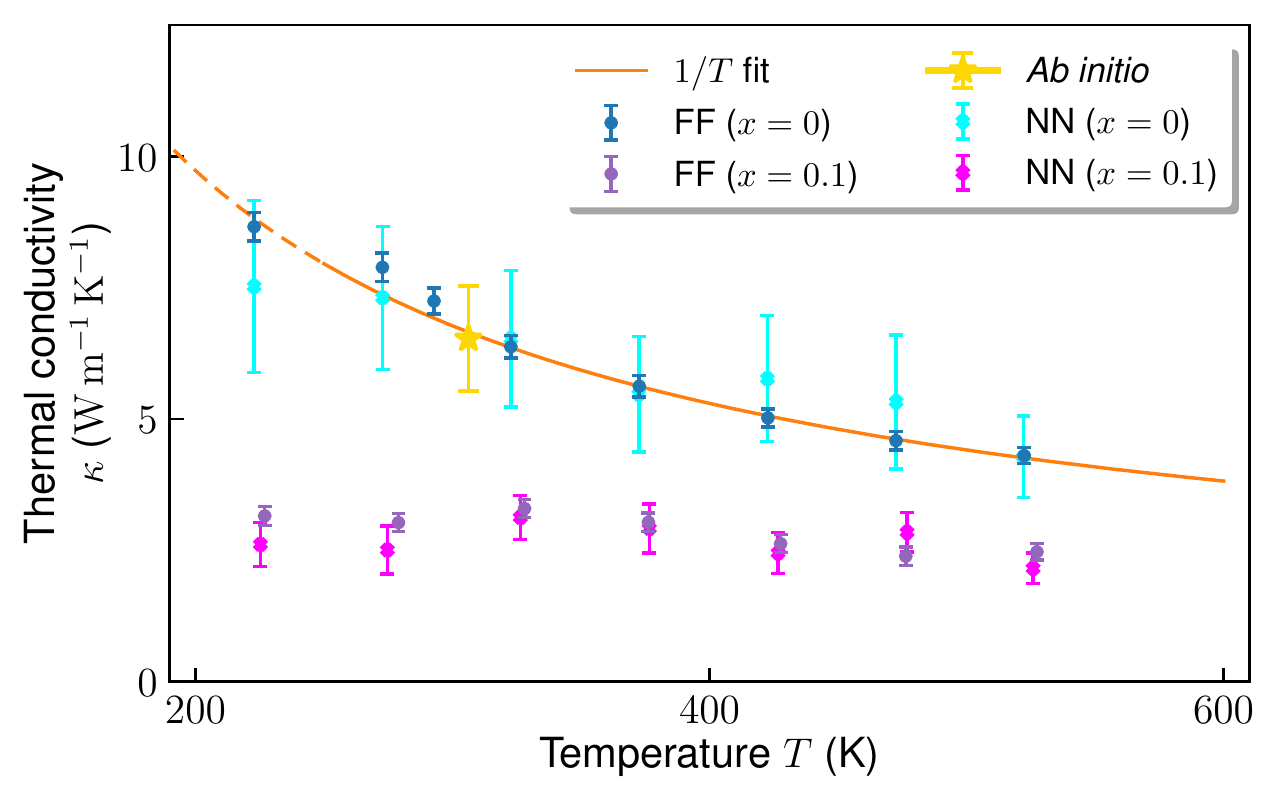}
    \caption{GK thermal conductivity of Li\textsubscript{3}ClO. Results are obtained from MD simulations with the classical FF and with the NN model for the perfect crystalline system ($x=0$) and the highest concentration $(x=0.1)$ of vacancies we investigate. The $1/T$ fit of the high temperature behaviour of the FF data at $x=0$ is shown in orange. The yellow star is the \emph{ab initio} GK result at RT. Error bars represent standard deviations.}
    \label{fig: thermal conductivity FF vs NN}
\end{figure}

As for the machine learning model, a DeepPot-SE NN~\cite{han2017deep, zhang2018end} is trained on a $3\times 3 \times 3$ supercell of Li\textsubscript{3}ClO with a LiCl pair removed. Details on the model and its validation can be found in Appendix~\ref{app: NN}. NN based MD simulations are carried out for $x=0$ and $x=0.1$ across the whole temperature range of interest. Thermal conductivity for the NN model is computed using the methodology developed in Ref.~\onlinecite{tisi2021heat}. In Fig.~\ref{fig: thermal conductivity FF vs NN} we show a comparison between the GK thermal conductivity obtained with classical-FF- and the NN-potentials for these two systems. Results from NN simulations are in close agreement with the AI ones available at room temperature and, remarkably, with those obtained from classical FFs, over a broad temperature range, thus further validating the accuracy of the latter for the purposes of the present work.

Having thoroughly verified that the classical FF closely mimics AI-quality results, we use it to perform a systematic analysis of the thermal transport properties of Li\textsubscript{3}ClO in a broad range of temperatures and vacancy concentrations.
MD simulations are carried out on ${10 \times 10 \times 10}$ supercells. The desired vacancy concentration is obtained by randomly removing the corresponding number of $\mathrm{LiCl}$ pairs from the supercell by means of the \texttt{Atomsk} code.~\cite{hirel2015atomsk} For each temperature, we employed the temperature-dependent \emph{ab initio} lattice parameter computed in Sec.~\ref{sec:SP} in the QHA (independently of vacancy concentration). In the equilibration phase, the canonical ensemble is sampled via the CSVR thermostat~\cite{bussi2007canonical} for $200\,\mathrm{ps}$. At this point, the thermostat is removed and a microcanonical ($NVE$) production run of $5\,\mathrm{ns}$ is carried out to collect the desired data.

\begin{figure}
    \centering
    \includegraphics[width=\columnwidth]{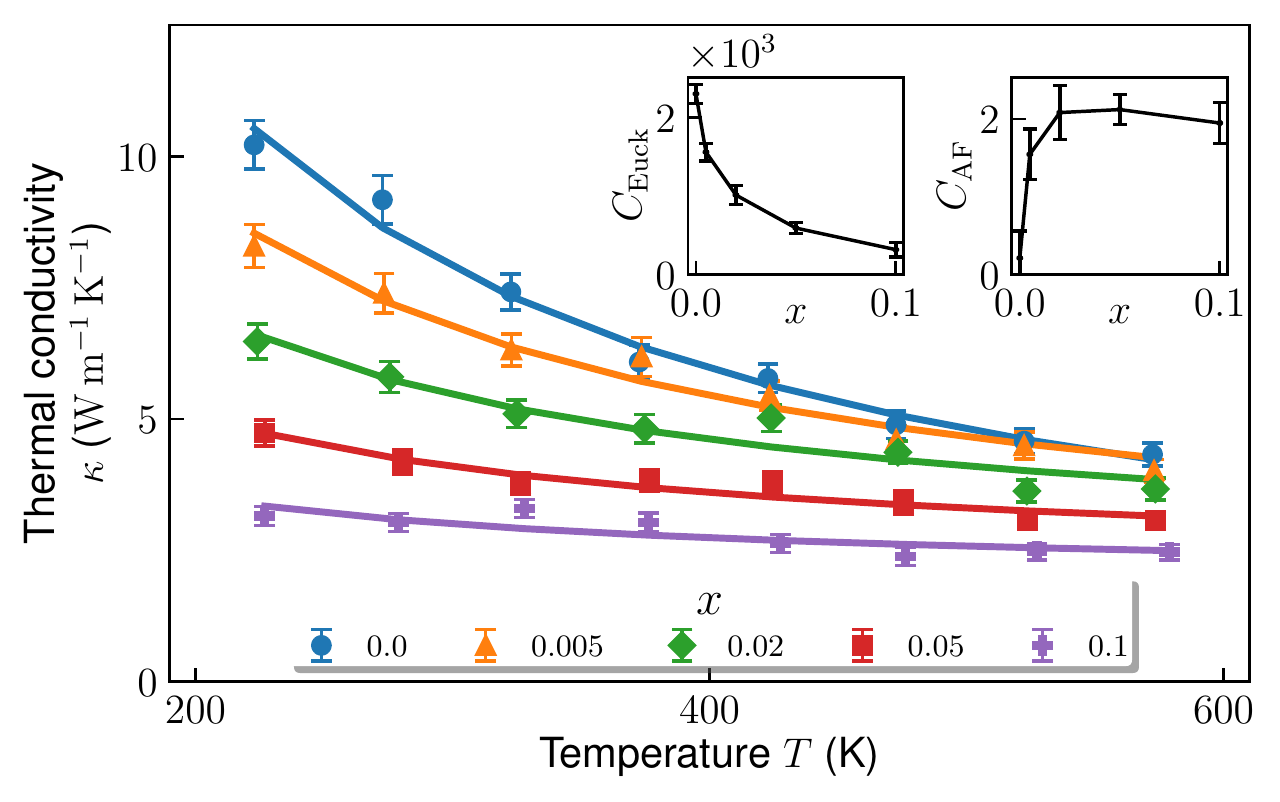}
    \caption{GK thermal conductivity as a function of temperature for different values of the concentration of vacancies, $x$. The solid lines are fits performed according to Eq.~\eqref{eq: kappa fitting function}. The insets show the fitting parameters, $C_{\mathrm{Euck}}$ and $C_{\mathrm{AF}}$, in units of $\mathrm{W\,m^{-1}}$ and $\mathrm{W\,m^{-1}\,K^{-1}}$, respectively, as a function of $x$. The fitting parameters are also reported in Table~\ref{tab: fit}. Error bars represent standard deviations. }
    \label{fig: thermal conductivity vs vacancies}
\end{figure}

\begin{figure*}[tb]
    \centering
    \includegraphics[width=2\columnwidth]{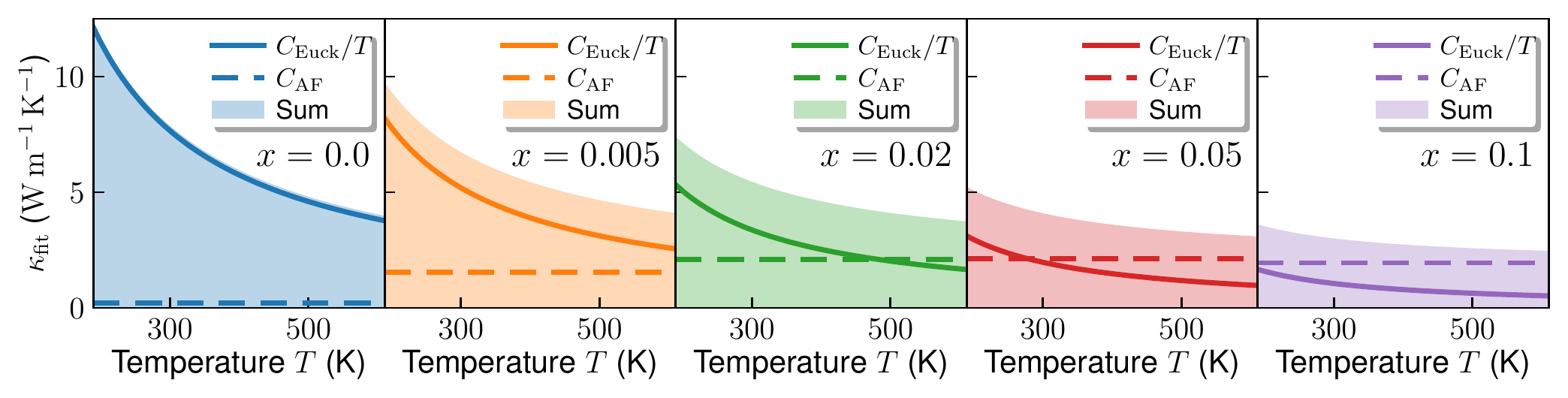}
    \caption{Eucken and Allen-Feldman contributions to the fit of Eq.~\eqref{eq: kappa fitting function} on the GK thermal conductivity data.}
    \label{fig: GK fit contributions}
\end{figure*}

The dependence of the thermal conductivity on $T$ and $x$ is shown in Fig.~\ref{fig: thermal conductivity vs vacancies}. The numerical data measured in the MD simulations are then fitted to a simple function, Eq.~\eqref{eq: kappa fitting function}, as described below. The functional form of such dependence should account for the Peierls-Boltzmann asymptotic $1/T$ behaviour (Eucken's law),~\cite{eucken1911temperaturabhangigkeit} holding for crystals ($x=0$) at large temperatures, but should also be able to take into account the breakdown of crystalline order when vacancies are present ($x \neq 0$). The presence of vacancies, while providing access to Li-ion conduction channels, on shorter time-scales establishes an effective local disorder, that would result in a contribution to the thermal conductivity describable at different levels of accuracy~\cite{einstein1911elementare, allen1993thermal, isaeva2019modeling, simoncelli2019unified}. The simplest significant approach is to consider the Allen-Feldman model~\cite{allen1993thermal} of thermal conduction in harmonic glasses, where the thermal conductivity, $\kappa_{\mathrm{AF}}$, takes the form
\begin{align}\label{eq: allen feldman}
    \kappa_{\mathrm{AF}} = -\sum_{\lambda} \frac{\hbar \omega_{\lambda}^2}{\Omega^2 T} \pdv{n_\lambda(T)}{\omega_\lambda} D_{\lambda}.
\end{align}
Here, $\lambda$ is a label for the modes, $n_\lambda(T)$ is the Bose-Einstein occupation number at temperature $T$, and $D_\lambda$ is the (temperature-independent) mode diffusivity; asymptotic expansion of~\eqref{eq: allen feldman} in the high temperature limit yields
\begin{align}
    \kappa_{\mathrm{AF}} \sim \frac{k_\mathrm{B}}{\Omega^2}\sum_{\lambda} D_{\lambda} + \order{\frac{1}{T^2}},
\end{align}
\emph{i.e.}, the leading order is constant in temperature. Notice that MD with classical nuclei sample, by definition, classical distributions: the occupation of a mode of energy $\hbar \omega$ is thus the first order expansion of Bose-Einstein distribution, $n_\lambda(T) \approx k_\mathrm{B} T/\hbar \omega_\lambda$, and the heat capacity per mode reduces to the Dulong-Petit result, $C_\lambda \approx k_\mathrm{B}$. All in all, we take as fitting function for $\kappa(T,x)$
\begin{align}\label{eq: kappa fitting function}
    \kappa_\mathrm{fit} = \frac{C_{\mathrm{Euck}}}{T} + C_{\mathrm{AF}},
\end{align}
where $C_{\mathrm{Euck}}(x)$ and $C_{\mathrm{AF}}(x)$ are vacancy-dependent fitting parameters, incorporating both the Eucken and the Allen-Feldman asymptotics. Their values are reported in Table~\ref{tab: fit} and shown in the inset of Fig.~\ref{fig: thermal conductivity vs vacancies}. As expected, $C_{\mathrm{AF}}$ is compatible with zero in the case $x=0$, while sensibly different in the other cases. It is also comforting to notice that $C_\mathrm{Euck}(x)$ vanishes as $x$ increases, while $C_\mathrm{AF}(x)$ saturates. To have a clear picture of the relative importance of these contributions, we explicitly report, in Fig.~\ref{fig: GK fit contributions}, the decomposition of $\kappa$ vs $T$ into the Eucken and AF terms at the different vacancy concentrations inspected in MD simulations. This picture suggests that, at least in this class of SSE, ionic diffusion may contribute very little to $\kappa$, which is instead dominated by the lattice component at low temperature, and by disorder at high temperature. This was recently investigated experimentally in Ref.~\onlinecite{bernges2021diffusons} for $\mathrm{Ag^{+}}$ fast-ion conductors.

Our GK calculations show that the thermal conductivity strongly depends on the presence of defects: in the considered temperature range, even a few-\% concentration of defects is able to almost halve the $\kappa$ with respect to its value for the perfect crystal. This behaviour is particularly evident at low temperature, due to the suppression of the Eucken law when $x$ increases. Nonetheless, when compared to other candidates for battery-oriented SSE, Li\textsubscript{3}ClO is characterised by a relatively high thermal conductivity, and meets the requirements for safe heat dissipation and management.

\section{Conclusions}

We conclude with a summary of our results and some perspectives on the possible use of Li\textsubscript{3}ClO in realistic devices. In the first part of our work we have reported state-of-the art calculations on the structural and mechanical properties of Li\textsubscript{3}ClO, including the zero-point lattice contribution and non-analytical corrections. According to our calculations, Li\textsubscript{3}ClO is mechanically stable, and characterised by a larger brittleness and anisotropy than previously thought. From the thermal transport standpoint, we extensively showed, by comparing different levels of theory, that current estimates of the thermal conductivity are fully unreliable, overestimating the value obtained via more accurate theory by several times in the temperature window at which SSE are supposed to operate. We explicitly demonstrated that, due to strong anharmonicity, not even the full solution of the Peierls-Boltzmann transport equation with three-phonon scattering is sufficient and higher order contributions must be considered, which tend to further decrease the thermal conductivity of Li\textsubscript{3}ClO. We performed molecular dynamics simulations and employed the GK theory of linear response and newly developed data-analysis tools to automatically incorporate all scattering orders and deal with vacancies and charge carrier diffusion as well. We observed that increasing the number of vacancies induces local disorder, which leads to a glass-like behaviour where the thermal conductivity saturates at large $T$ rather than vanishing like $1/T$ according to Eucken's law. Ionic diffusion \emph{per se} seems instead to affect  thermal transport only marginally. Our calculations indicate that, even if the capability of Li\textsubscript{3}ClO to dissipate heat in a realistic device must be scaled back, Li\textsubscript{3}ClO should still have a good thermal conductivity in a quite large temperature range and may be safely employed as a battery-oriented SSE. We hope that our work will encourage an experimental assessment of thermal transport properties of Li\textsubscript{3}ClO, and will also serve as a valid methodological reference, aimed at numerical-simulation practitioners, on how to investigate the thermal transport properties of generic SSE ionic conductors through state-of-the-art theoretical and data-analysis tools.

\section{Methods}

In this work, we follow a multi-method approach, hinging on a combination of classical simulations based on force fields (FF), \emph{ab initio} (AI) computations based on density-functional (perturbartion) theory, and neural-network (NN) modelling of \emph{ab initio} data. In all cases, the accuracy of empirical FFs and NN potentials is thoroughly benchmarked against AI data. 

Electronic, structural, mechanical, and vibrational properties are studied using DFT (Sec.~\ref{sec:meth:electronic structure}). The thermal conductivity is evaluated both through a Boltzmann-Peierls kinetic approach (Sec.~\ref{sec:BTE}) and via the GK theory of linear response (Sec.~\ref{sec: GK}) and classical MD. BTE calculations  are performed both \emph{ab initio}, and using classical FFs (Sec.~\ref{sec: classical FF}). AI data are used as a reference, while FF simulations allow us to estimate the magnitude of finite-size and anharmonic effects, whenever AI calculations would be unfeasible, as well as the role of vacancies in thermal transport in non-stoichiometric conditions.

\subsection{Electronic and atomic structure}\label{sec:meth:electronic structure}
All the electronic-structure calculations are performed within DFT and the plane-wave pseudopotential method using \textsc{Quantum~ESPRESSO}~\cite{qe1,qe2,qe3} and SG15 Optimized Norm-Conserving Vanderbilt~(ONCV) pseudopotentials.~\cite{oncv-pseudo} Structural optimization and lattice-dynamical calculations were performed using the generalised gradient approximation~(GGA) to the exchange-correlation functional in the Perdew–Burke–Ernzerhof~(PBE) flavour,~\cite{perdew1996generalized} while the electronic band structure is computed using the Heyd-Scuseria-Ernzerhof~(HSE06)~\cite{heyd2003hybrid} hybrid functional. The electronic structure is obtained starting from a self-consistent calculation {on a coarse $\mathbf k$-point grid in the Brillouin zone (BZ) leveraging} a fast implementation of the Fock-exchange operator,~\cite{carnimeo2019fast} followed by a transformation to the Wannier basis, performed with \textsc{Wannier90}.~\cite{pizzi2020wannier90} and a subsequent non-self-consistent calculation to interpolate the energy bands. All calculations are performed using a plane-wave kinetic energy cutoff of $200\,\mathrm{Ry}$ (such an unusually high cutoff was needed for an optimal convergence of equilibrium thermal properties, \emph{vide infra}) and the Brillouin Zone~(BZ) sampling is done via a Monkhorst-Pack grid~\cite{monkhorst1976special} of ${12\times 12 \times 12}$ points displaced by half a grid step along the $(1,1,1)$ direction. Vibrational properties are computed via Density Functional Perturbation Theory~(DFPT).~\cite{baroni2001phonons} dynamical matrices have been calculated on a ${7 \times 7 \times 7}$ $\mathbf{q}$-point grid, and Fourier-interpolated on a ${192 \times 192 \times 192}$ mesh to evaluate the temperature dependence of the free energy in the quasi-harmonic approximation,~\cite{Baroni2010QHA} with the aid of the \texttt{thermo\_pw} \textsc{Quantum~ESPRESSO} driver.~\cite{malica2020quasi}

\subsection{Classical molecular dynamics}\label{sec: classical FF}

Molecular Dynamics~(MD) simulations have been carried out with \textsc{lammps},~\cite{lammps, lammps2} using different kinds of force fields. Specifically, we adopt both an empirical FF and a NN interatomic potential trained and validated on \textit{ab initio} data. We anticipate that the overall very good agreement between FF and NN simulations safely justifies the extended use of the empirical FF to study thermal transport across a wide range of temperatures and vacancy concentrations in a computationally affordable way.

The empirical force-field, which was fitted to structural and dielectric properties from experiments and \emph{ab initio} calculations, is taken from Ref.~\onlinecite{mouta2014concentration}. We adopt the simplified version based on a rigid-ion model, as opposed to core-shell models that are necessary when addressing specifically the electronic polarisability. The rigid-ion approximation has been used by other authors to investigate charge transport.~\cite{sagotra2017stress, lu2015defect} The short-range interatomic interaction is modelled by a Buckingham potential of the form
\begin{align}\label{eq: buckingham}
    \phi_{ij} = A_{ij} \mathrm{e}^{
    -\frac{r_{ij}}{\rho_{ij}}}
    - \frac{C_{ij}}{r^6_{ij}},
\end{align}
where $i$ and $j$ label the two atoms separated by a distance $r_{ij}$, while $A_{ij}$, $\rho_{ij}$ and $C_{ij}$ are fitting parameters. The long-range (Coulomb) interaction among point charges is computed via the Ewald summation technique.~\cite{ewald1921berechnung} The time-step for the dynamics is of $1\,\mathrm{fs}$.

A Deep Potential--Smooth Edition (DeepPot-SE) NN model~\cite{han2017deep, zhang2018end, wang2018deepmd} is employed to accurately mimic AI atomic forces in a computationally affordable way. The model is trained on a set of configurations sampled from Car-Parrinello simulations~\cite{car1985unified} enriched with additional configurations chosen by the \texttt{dpgen} software.~\cite{dpgen} Further details are provided in Appendix~\ref{app: NN}.

\subsection{Boltzmann transport equation}\label{sec:BTE}

The BTE is solved by real-space techniques with
the \texttt{ShengBTE} code~\cite{li2014shengbte} and its successor, \texttt{almaBTE},~\cite{carrete2017almabte} using harmonic and higher-order interatomic force constants (IFC) from both classical FFs, so as to make a comparison with GK-MD results handy, and first-principles methods based on DFT. In the first case, IFCs are obtained in real space using \texttt{phonopy}.~\cite{phonopy} Details on size effects and the choice of the parameters can be found in Sections~\ref{ssec:IFC3-sc},~\ref{ssec:IFC3-nnn} and~\ref{ssec:IFC3-nq} of Appendix~\ref{app: BTE}. Quadratic and cubic force constants AI IFCs are obtained from second- and third-order DFPT, using the \texttt{ph.x} and \texttt{D3Q}~\cite{paulatto2013anharmonic, fugallo2013ab} components of \textsc{Quantum~ESPRESSO}, respectively.

\subsection{Green-Kubo linear-response theory}\label{sec: GK}

In the linear-response regime, the thermal conductivity, $\kappa$, is defined as the ratio between the energy flux and the negative of the temperature gradient \textit{in the absence of any convection}. For solids and one-component fluids this last prescription is trivially satisfied in practical MD simulations performed in the barycentric reference frame. Within the GK theory of linear response, the thermal conductivity is given by the celebrated GK formula~\cite{green1952markoff, green1954markoff, kubo1957statistical1, kubo1957statistical2, baroni2020heat}
\begin{align}\label{eq: GK one component}
    \kappa \propto \int_0^\infty \expval{\mathbf{J}_E(t) \cdot \mathbf{J}_E(0)} \dd{t},
\end{align}
where the energy flux, $\mathbf{J}_E(t)$, reads
\begin{align}\label{eq: J_E}
    \mathbf{J}_E(t) = \frac{1}{\Omega} \sum_i \left[ \epsilon_i \mathbf{V}_i + \sum_{j} (\mathbf{F}_{ji} \cdot \mathbf{V}_i) (\mathbf{R}_i-\mathbf{R}_j) \right].
\end{align}
Here $\Omega$ is the system's volume, $\epsilon_i$ is the energy assigned to the $i$-th atom, $\mathbf{V}_i$ and $\mathbf{R}_i$ its velocity and position, respectively, and $\mathbf{F}_{ji}=-\frac{\partial \epsilon_i}{\partial \mathbf R_j}$.

For a multi-component system such as the superionic phase of Li\textsubscript{3}ClO, Eq.~\eqref{eq: GK one component} cannot be applied as it is, since the prescription of vanishing convection is not automatically satisfied in a standard MD simulation: in fact, keeping the barycentre of the whole system fixed does not imply that the barycentres of each atomic species $S$ (with ${S=\mathrm{Li}, \mathrm{Cl}, \mathrm{O}}$) stay fixed. Therefore, the mass fluxes of two of the three species, say $\mathbf{J}_\text{Li} (t) =\frac{M_\text{Li}}{\Omega} \sum_{i\in \text{Li}} \mathbf V_i(t)$ and $\mathbf{J}_\text{Cl} (t) =\frac{M_\text{Cl}}{\Omega} \sum_{i\in \text{Cl}} \mathbf V_i(t)$, where $M_S$ is the atomic mass of species $S$, \emph{i.e.} the total momenta of each atomic species, are independent and in general non vanishing.  Notice that, from a more fundamental point of view, even the concept of \emph{atomic energy} $\epsilon_i$ appearing in Eq.~\eqref{eq: J_E} is intrinsically flawed, in that \textit{i)} there is no \emph{a priori} ``correct'' decomposition of the total energy of a system of interacting atoms among its constituents, and \emph{ii)} the zero of the atomic energies (\emph{i.e.} the energy of isolated atoms) is arbitrary. A theoretically sound solution to both these problems has been provided in \nocite{marcolongo2016microscopic,ercole2016gauge,baroni2020heat} Refs.~ \onlinecite{marcolongo2016microscopic}, \onlinecite{baroni2020heat}, and \onlinecite{ercole2016gauge}, where it is rigorously proved that these apparent inconsistencies \textit{must} disappear when a \textit{measurable} quantity, such as $\kappa$, is correctly calculated, leading to the formulation of so-called invariance principles of transport coefficients.~\cite{grasselli2021invariance,marcolongo2020gauge} Furthermore, a multivariate  technique~\cite{ercole2017accurate, bertossa2019theory} for the analysis of the energy flux time-series obtained from MD simulations has been developed, which allows one to compute $\kappa$ for multicomponent systems, like the superionic Li\textsubscript{3}ClO, in an efficient and rigorous way. We redirect the reader to Refs.~\onlinecite{ercole2017accurate} and~\onlinecite{bertossa2019theory} for a thorough description of the method. Suffice it here to say that the thermal conductivity for Li\textsubscript{3}ClO is estimated in terms the $\omega \to 0$ limit of the power spectrum of the energy flux, after removing its coupling to the mass fluxes. This is achieved by computing the Schur complement of the matrix whose entries are the power spectra between fluxes $\mathbf{J}_A$, $\mathbf{J}_B$, i.e. ${S_{AB}(\omega) = \int_{-\infty}^{\infty} e^{i\omega t} \langle \mathbf J_A(t) \cdot \mathbf J_B (0) \rangle \dd{t}}$, where $A,B\in\{E, \mathrm{Li}, \mathrm{Cl}\}$ refer to energy flux, Li mass flux, and Cl mass flux, respectively:
\begin{align}\label{eq: kappa_multi}
    \begin{split}
        \kappa(\omega) &= \frac{\Omega}{6k_\mathrm{B} T^2} \left[
        S_{EE}(\omega) - S_{\text{coupl.}}(\omega) \right] \\
        S_{\text{coupl.}}&\equiv
        \begin{pmatrix}
        S_{\text{Li } E} & S_{\text{Cl } E}
        \end{pmatrix}
        \begin{pmatrix}
        S_{\text{Li Li}} & S_{\text{Li Cl}} \\
        S_{\text{Cl Li}} & S_{\text{Cl Cl}}
        \end{pmatrix}^{-1}
        \begin{pmatrix}
        S_{E \text{ Li}} \\
        S_{E \text{ Cl}}
        \end{pmatrix}
    \end{split}
\end{align}
Since Li\textsubscript{3}ClO is a three-components material, two of the three mass fluxes are independent (the third being fixed by conservation of total momentum) and as such must, in general, be decoupled from the energy flux. In the case where some mass component were non-diffusive, they would contribute nothing to the value of the decoupled spectrum at zero frequency. Nonetheless, it was shown numerically that the decoupling at finite frequency allows a better estimate of the value at ${\omega=0}$.~\cite{bertossa2019theory}. The numerical tools needed to evaluate the power spectra of time series of $\mathbf J_{E}$, and $\mathbf J_{\text{Li}}$ and $\mathbf{J}_{\text{Cl}}$ have been implemented in the open-source \textsc{SporTran} code,~\cite{sportran} which we extensively adopted in the present work.

\section*{Data availability}
Numerical data supporting the plots and relevant results within this paper, the neural network potential and the data set used to generate it are available on the Materials Cloud Platform~\cite{talirz2020materials}.

\section*{Acknowledgements}
This work was partially funded by the EU through the \textsc{MaX} Centre of Excellence for supercomputing applications (Project No. 824143) and by the Italian Ministry of Research and education through the PRIN 2017 \emph{FERMAT} grant. FG acknowledges funding from the Swiss National Science Foundation (SNSF), through Project No.~200021-182057, and from the European Union's Horizon 2020 research and innovation programme under the Marie Sk\l{}odowska-Curie Action IF-EF-ST, grant agreement No.~101018557 (TRANQUIL). We thank Cristiano Malica for his support with \emph{ab initio} thermodynamics methods; Davide Tisi and Cesare Malosso for their support with machine learning methods; Pietro Delugas, Riccardo Bertossa and Alfredo Fiorentino for technical support and useful discussions; Kevin Rossi for a critical reading of the manuscript.





\appendix 

\section{Slack model}\label{app: slack}

For a crystal with $n$ atoms per unit cell, whose average atomic mass is $\overline{M}$, thermal conductivity in $\mathrm{Wm^{-1}K^{-1}}$ is estimated as
\begin{align}
    \kappa = \frac{2.43 \cdot 10^{-6}}{1 - 0.514/\gamma + 0.228/\gamma^2} \frac{\overline{M} \Theta_D^3 \delta}{\gamma^2 T n^{2/3}},
\end{align}
where $\Theta_D$ is the Debye temperature in $\mathrm{K}$, $\gamma$ is the Grüneisen parameter, $\delta$ is the cubic root of the average atomic volume in \AA, and $\overline{M}$ is expressed in atomic mass units. The Debye temperature is computed directly from the Debye model, by evaluating the average sound velocity obtained from the angular average of the sound velocities, which are in turn calculated solving the wave equation for each propagation direction; the other quantities are described in the main text.

\section{Boltzmann transport equation}\label{app: BTE}

\begin{figure*}[htb]
    \centering
    \includegraphics[width=2\columnwidth]{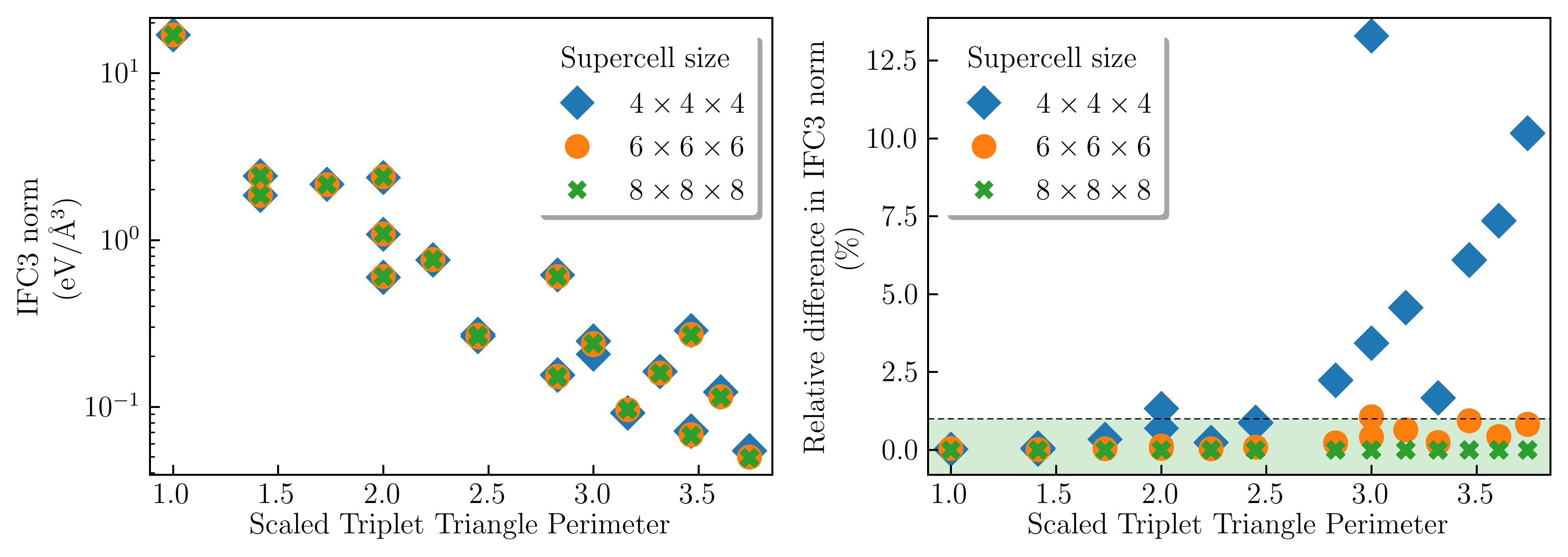}
    \caption{Convergence of the IFC3 with respect to the number of supercell size. (\emph{left}) Triangle plots for different supercell sizes. (\emph{right}) Relative difference in the IFC3 norms with respect to the reference one with a ${8\times8\times8}$ supercell. The green-shaded area indicates values below $1\%$.}
    \label{fig: triangle plot supercell two}
\end{figure*}

Following Barbalinardo \emph{et al.},~\cite{barbalinardo2020efficient} let us label the phononic states of a general anisotropic crystal at a finite temperature $T$ by a wave-vector $\bm{q}$ and a band index $s$. Phonons populate the available energy levels according to the Bose-Einstein statistic with null chemical potential, \emph{i.e.}
\begin{align}\label{eq: BE occupation number}
    n_{\bm{q} s} = n(\omega
    _{\bm{q} s}) = \frac{1}{\mathrm{e}^{\nicefrac{\hbar\omega_{\bm{q} s}}{k_\mathrm{B} T}} - 1},
\end{align}
with $\omega_{\bm{q} s}$ the angular frequency of vibration of the normal mode labeled by $(\bm{q}, s)$. Under the assumption that the phonon population depends on position only through the temperature, in linear response to a small temperature gradient $\nabla_\alpha T$ along direction $\alpha=x,y,z$, the occupation number~\eqref{eq: BE occupation number} acquires a contribution proportional to $\nabla_\alpha T$:
\begin{align}\label{eq: occupation number Taylor expanded}
    \tilde{n}_{\bm{q} s \alpha} = n_{\bm{q} s} + \lambda_{\bm{q} s} \nabla_\alpha n_{\bm{q} s} = n_{\bm{q} s} + \lambda_{\bm{q} s} \pdv{n_{\bm{q} s}}{T} \nabla_\alpha T.
\end{align}
The quantity $\lambda_{\bm{q} s}$ is called phonon mean free path. From the $\bm{q}$-gradient of the phonon frequencies, one obtains the phonon group velocities $v_{\bm{q} s \alpha}=\pdv{\omega_{\bm{q} s}}{q_\alpha}$; these, together with the phonon energies $\hbar \omega_{\bm{q} s}$ and the out-of-equilibrium occupation numbers $\tilde{n}_{\bm{q} s \alpha}$, are used to define the heat current per normal mode as
\begin{align}\label{eq: heat current BTE}
    \begin{split}
        j_{\bm{q} s \alpha'} &= \sum_\alpha \hbar \omega_{\bm{q} s} v_{\bm{q} s \alpha} (\tilde{n}_{\bm{q} s \alpha} - n_{\bm{q} s}) \\
        &\simeq - \sum_\alpha \hbar \omega_{\bm{q} s} v_{\bm{q} s \alpha} \pdv{n_{\bm{q} s}}{T} \lambda_{\bm{q} s} \nabla_\alpha T.
    \end{split}
\end{align}
An expression for the thermal conductivity is retrieved after having introduced the Fourier equation of diffusive conduction, ${J_\alpha = -\sum_\alpha' \kappa_{\alpha\alpha'} \nabla_{\alpha'}T}$, that relates the macroscopic heat flux ${J_\alpha=\sum_{\bm{q}, s} j_{\bm{q} s \alpha} / (N \Omega)}$ to the temperature gradient. Thermal conductivity reads
\begin{align}
    \kappa_{\alpha \alpha'} = \frac{1}{\Omega N} \sum_{\bm{q}, s} \hbar  \omega_{\bm{q} s} \pdv{n_{\bm{q} s}}{T} v_{\bm{q} s \alpha} \lambda_{\bm{q} s}.
\end{align}
Thus, the computation of $\kappa$ requires the knowledge of the out-of-equilibrium occupation number. The statistical mechanics of an out-of-equilibrium system can be described by BTE, whose expression
is
\begin{align}\label{eq: BTE}
    \bm{v}_{\bm{q} s} \cdot \nabla T \pdv{n_{\bm{q} s}}{T} = \left. \pdv{n_{\bm{q} s}}{t} \right|_{\text{scatt.}}.
\end{align}
The right-hand side is the \emph{scattering term}, whose linearized form contains all the scattering rates relating three-phonons scattering events. The inverse of the matrix of scattering rates is in turn used to compute the phonon mean free paths, which are what is needed to obtain $\kappa$.
The problem of computing $\kappa$ is reduced to the calculation and inversion of the scattering rates matrix. This can be demanding, and often requires additional approximations, the simplest of which being the relaxation time approximation~(RTA). In RTA, only the diagonal part of the scattering rates matrix is retained: the population of each out-of-equilibrium mode is taken to be interacting with a bath of modes at thermal equilibrium. A way to go beyond this picture is to recast the inversion problem to an iterative equation to be solved self-consistently.~\cite{omini1996beyond}

\subsection{Convergence of the IFC3 with the supercell}\label{ssec:IFC3-sc}

The Interatomic Force Constant (IFC) matrices are computed via finite differences on a supercell. The second order IFC (IFC2) is calculated inexpensively on a ${6\times6\times6}$ supercell. The choice of the supercell for the third order IFC (IFC3) changes substantially the computational cost of the calculations, and requires more care. 
The IFC3 is computed as a function of the supercell size. This relationship is studied via the so-called ``triangle plot''. The triangle plot for the IFC3 computed with 2, 4, 6 and 8 unit cells in each dimension is shown in Fig.~\ref{fig: triangle plot supercell two}. In the horizontal axis there is the perimeter of the triangle whose vertices are positions of the three atoms involved in each block in the IFC3 matrix, a block being the ${3  \times 3  \times 3}$ tensor $\frac{\partial^3{U}}{\partial\bm{r}_i \partial\bm{r}_y \partial\bm{r}_k}$. In the vertical axis there are the matrix norms of each block. A supercell of size ${6\times6\times6}$ is found to be sufficient to have a IFC3 converged up to $\approx 1\%$.

\subsection{Convergence of the IFC with the number of nearest neighbour shells}\label{ssec:IFC3-nnn}

\begin{figure*}[ht]
    \centering
    \includegraphics[width=2\columnwidth]{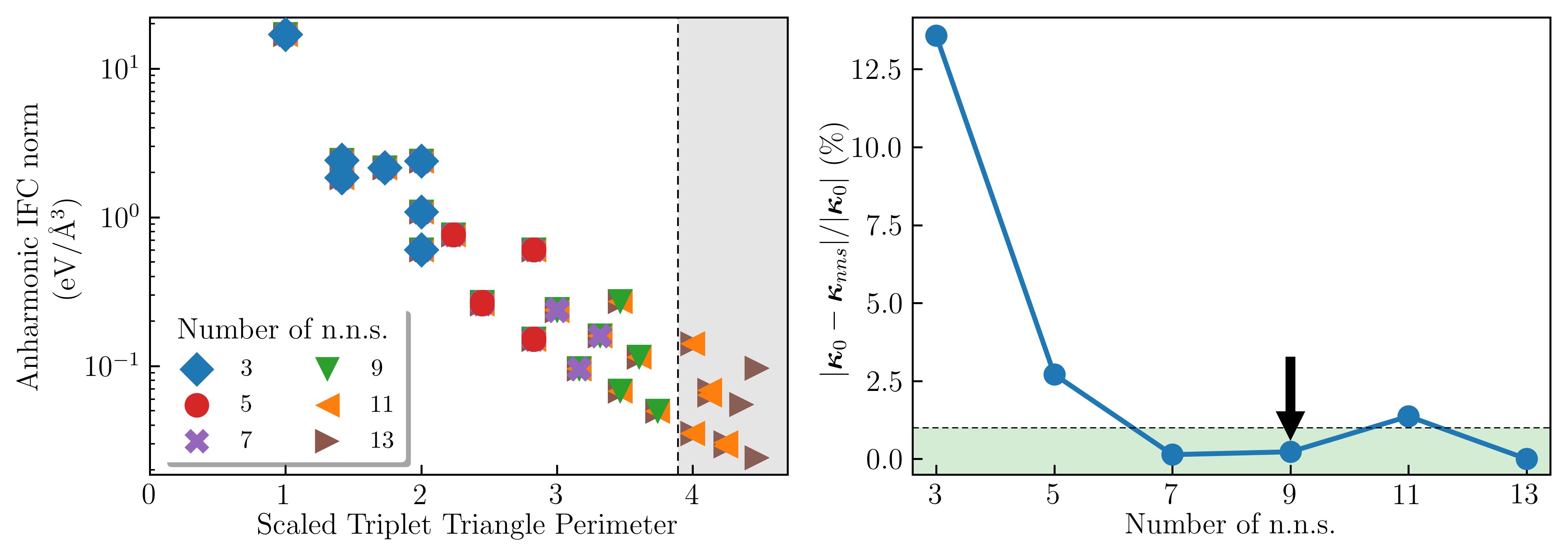}
    \caption{Convergence with respect to the number of n.n.~shells. (\emph{left}) Triangle plots for diffent values of the number of n.n. shells; (\emph{right}) relative difference (in Frobenius norm sense) in the lattice thermal conductivity from IFC3 computed with different n.n.~shells. The green-shaded area indicates values below $1\%$.}
    \label{fig: triangle plot NN and convergence of kappa}
\end{figure*}

The dependence on the number of included nearest neighbour (n.n.)~shells of FC3 is studied both via the triangle plot and by directly computing the thermal conductivity tensor, $\bm{\kappa}$, on a coarse ${12\times12\times12}$ $\bm{q}$-points mesh. The triangle plot for the FC3 computed with ${6 \times 6 \times 6}$ supercell is shown in the left panel of Fig.~\ref{fig: triangle plot NN and convergence of kappa}, while on the right panel there is the relative change in $\bm{\kappa}$ for a given number of n.n.~shells, denoted by $\bm{\kappa}_{nns}$, with respect to $\bm{\kappa}_0$, i.e.~the thermal conductivity tensor obtained with the highest number of n.n.~shells available. Frobenius matrix norms are employed, to take into account the off-diagonal elements as well. A number of n.n.~shells equal to $9$ is found to be sufficient to have a thermal conductivity tensor converged well below $1\%$.

\subsection{Convergence of the thermal conductivity with the number of $\bm{q}$-points}\label{ssec:IFC3-nq}

\begin{figure}[b]
    \centering
    \includegraphics[width=\columnwidth]{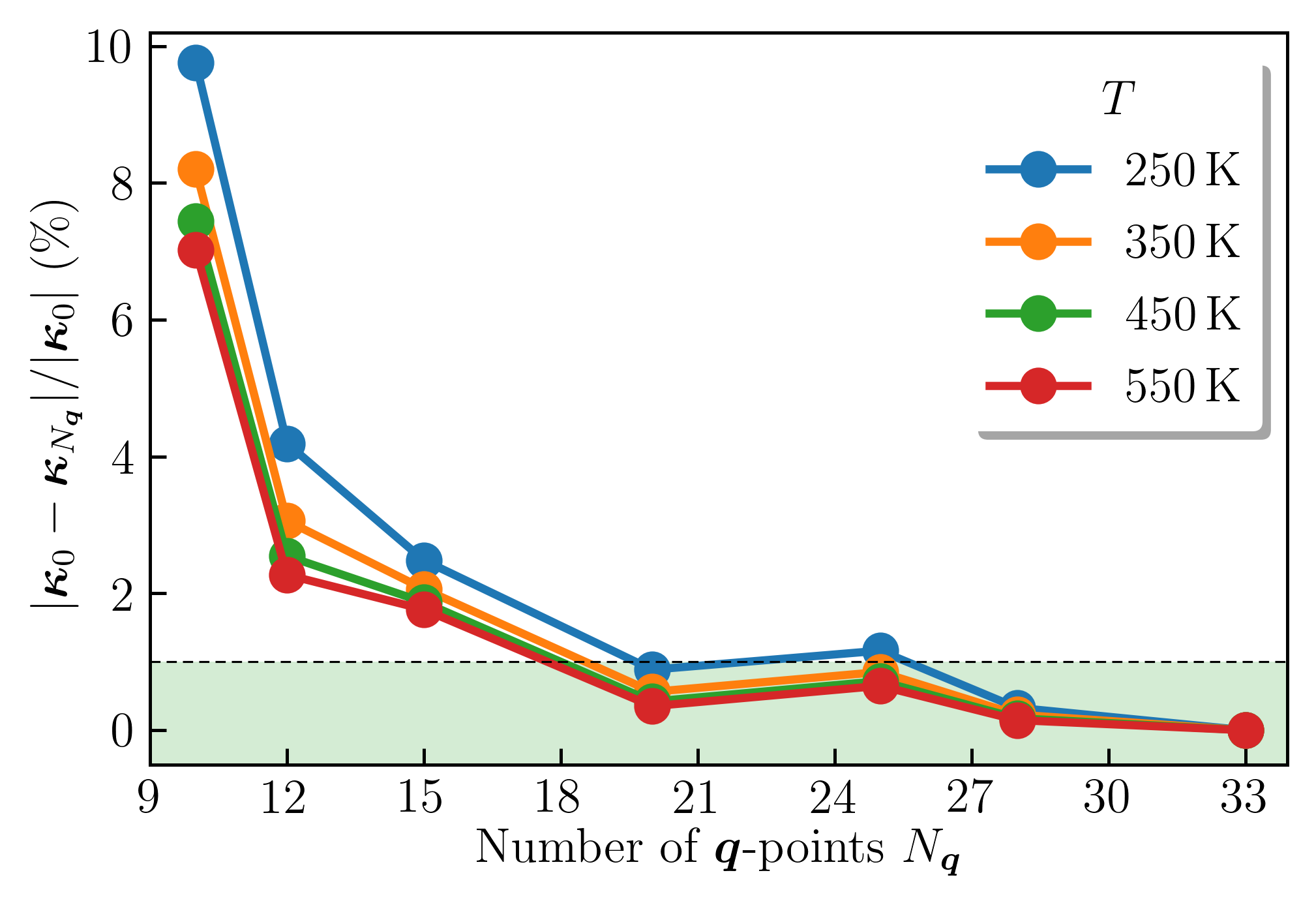}
    \caption{Convergence with respect to the number of $\bm{q}$-points.}
    \label{fig: q-mesh convergence}
\end{figure}

The convergence of $\kappa$ with the number of $N_{\bm{q}}$ of $\bm{q}$-points is analysed by computing the thermal conductivity tensor for different reciprocal space meshes for different temperatures. The results are then analysed in terms of the relative change in $\bm{\kappa}$ for a given $N_{\bm{q}}$, denoted by $\bm{\kappa}_{N_{\bm{q}}}$, with respect to $\bm{\kappa}_0$, i.e. the thermal conductivity tensor obtained with the highest $N_{\bm{q}}$ available. Convergence is deemed as satisfactory when the change is consistently lower than $1\%$. From Fig.~\ref{fig: q-mesh convergence} one can see that a mesh of ${20\times20\times20}$ $\bm{q}$-points is sufficient to provide converged results.

\subsection{Relaxation time approximation vs. full solution of the BTE}\label{ssec:full-bte-vs-rta}

\begin{figure}[tbh]
    \centering
    \includegraphics[width=\columnwidth]{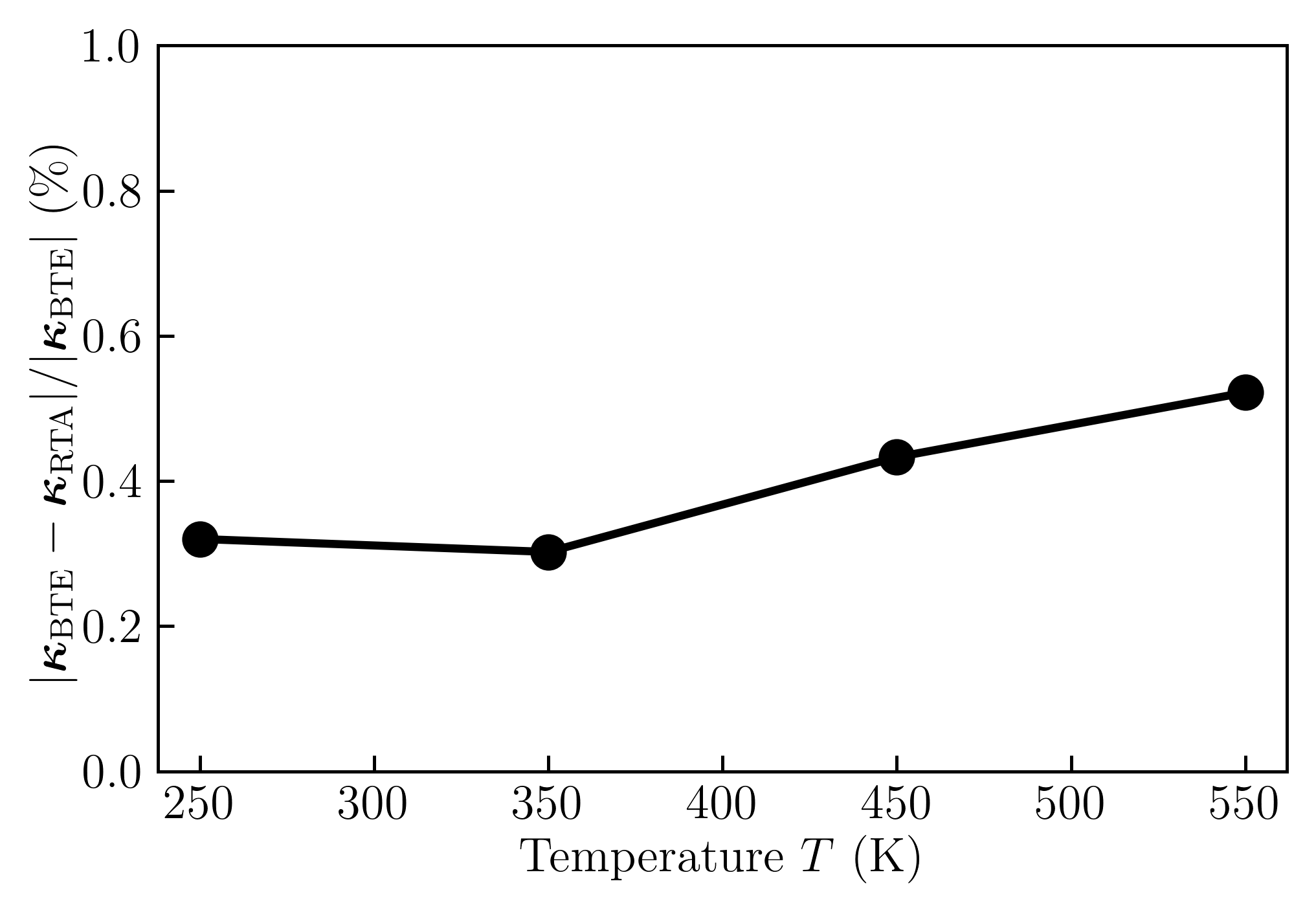}
    \caption{Relative difference between the RTA and the full solution to the linearised BTE.}
    \label{fig: rta vs bte}
\end{figure}

Calculations for selected temperatures are carried out to address the relative importance of the full solution of the linearised BTE with respect to the RTA result. As shown in Fig.~\ref{fig: rta vs bte}, it appears that RTA deviates from the full solution of at most $\approx 0.6\%$. Thus, it is employed for the rest of the calculations.

\subsection{Four-phonons contributions}\label{ssec:4-phonons}

Four-phonons calculations are carried out via the \texttt{Fourphonon} code~\cite{han2021fourphonon}, an extension to \texttt{ShengBTE}. Since four-phonon calculations are quite expensive, scattering rates and thermal conductivity for three- and four-phonon processes are performed on a coarse mesh of ${10\times10\times10}$ $\bm{q}$-points with IFCs coming from a ${4\times 4\times 4}$ supercell, and with a fourth-order interatomic force constant matrix (IFC4) that includes contributions from the first n.n. shell only. The scale parameter for Gaussian broadening is reduced to $0.1$ from the default value of $1.0$.

\subsection{Effect of the inclusion of NAC on thermal conductivity}\label{ssec:NAC}

\begin{figure}[tbh]
    \centering
    \includegraphics[width=\columnwidth]{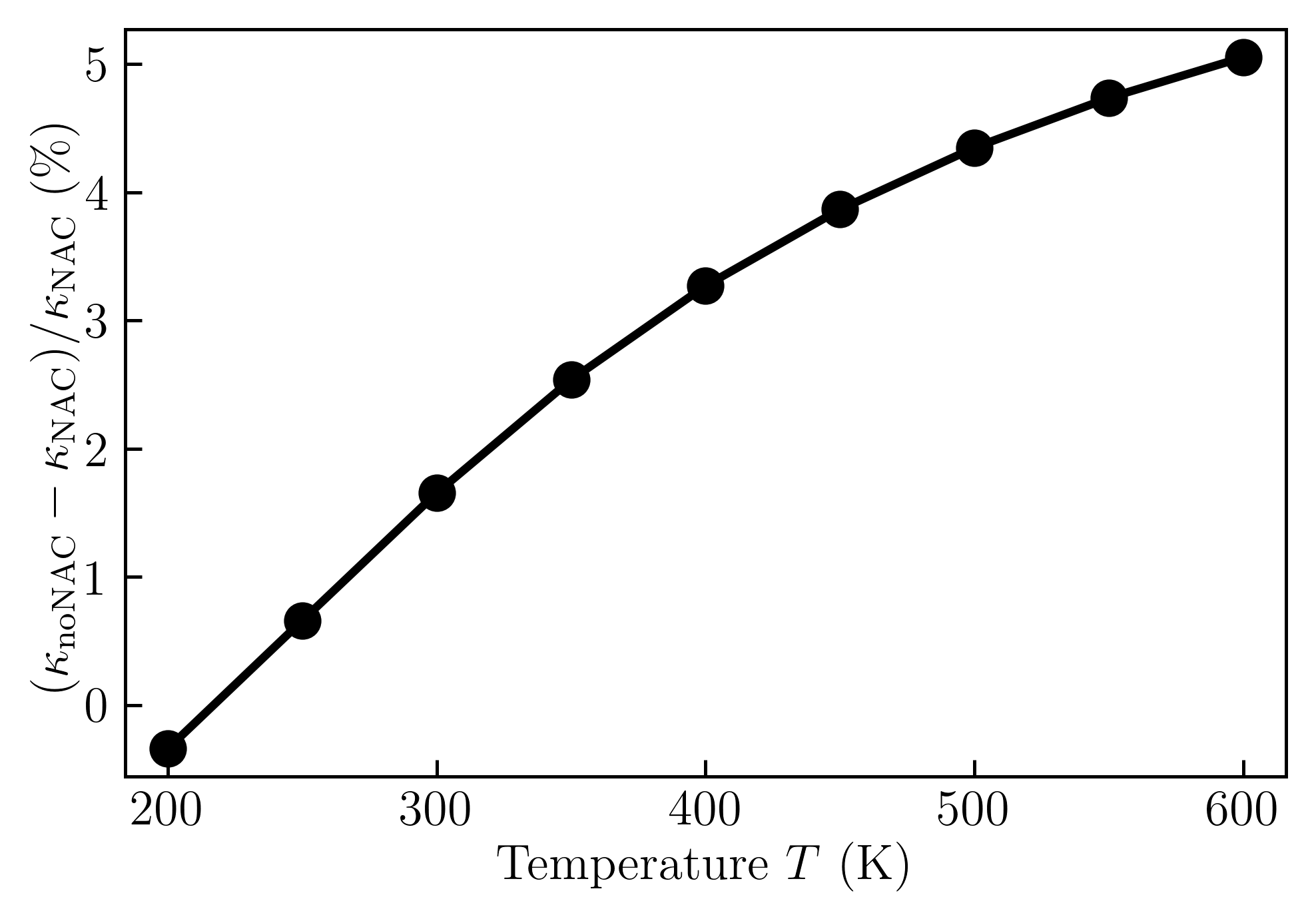}
    \caption{Relative difference in thermal conductivity including versus excluding NAC.}
    \label{fig: NAC vs no NAC}
\end{figure}

The effect on $\kappa$ of the inclusion/exclusion of NAC to the dynamical matrix are addressed in Fig.~\ref{fig: NAC vs no NAC}. The exclusion of NAC is found to produce a variation of up to $5\%$ to the value of $\kappa$ in the temperature range of interest.

\subsection{Equipartition vs. Bose-Einstein statistics}\label{ssec:equipartition-vs-be}

\begin{figure}
    \centering
    \includegraphics[width=\columnwidth]{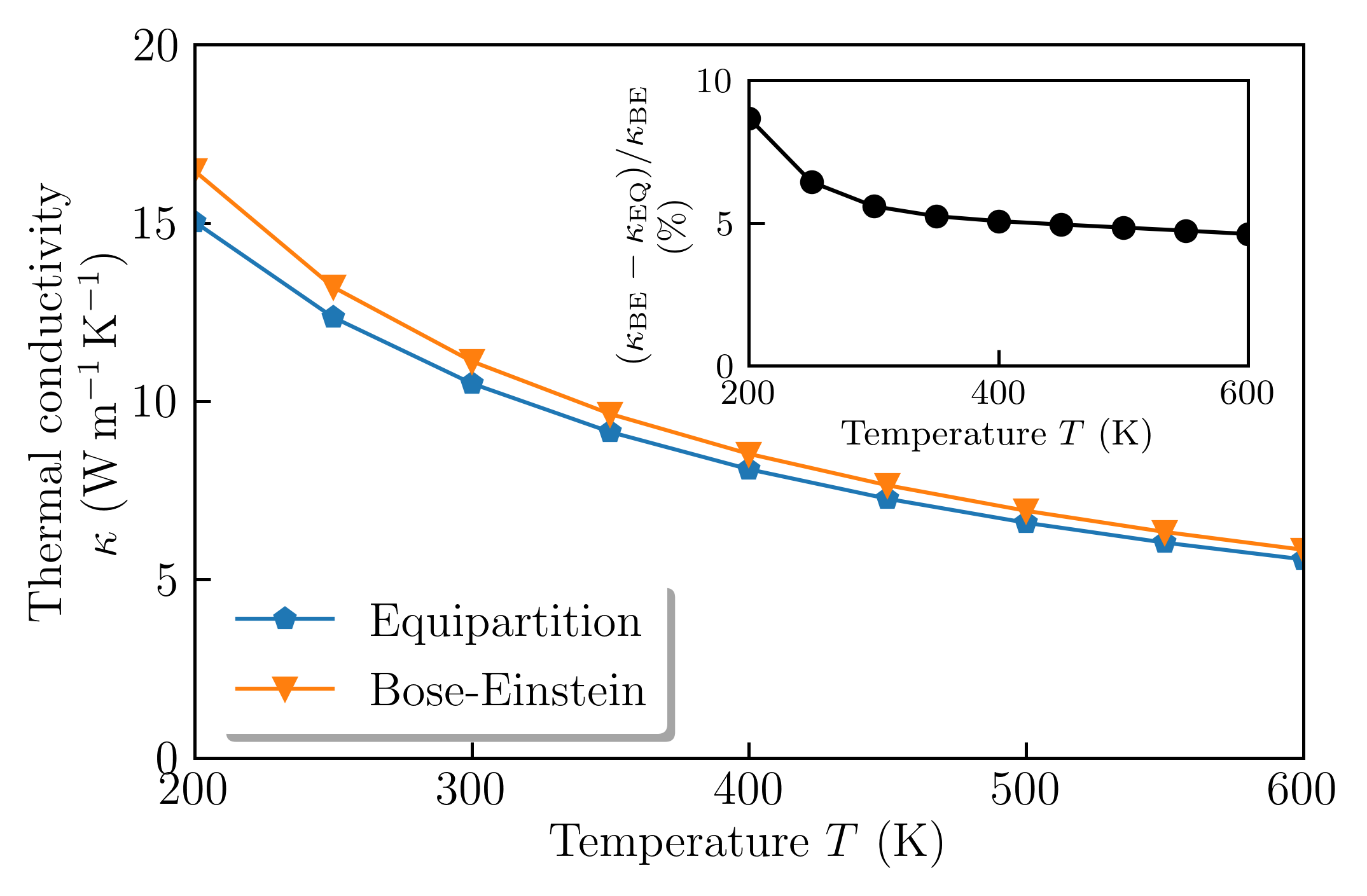}
    \caption{Comparison between lattice thermal conductivity computed with Bose-Einstein occupations and with the equipartition law. In the inset, their relative difference is plotted as a function of temperature.}
    \label{fig: BE vs EQ}
\end{figure}

In a BTE approach one can choose the statitstics of the phonon occupation numbers. For quantum particles, it is the Bose-Einstein distibution
\begin{align}\label{eq: BE}
    n_{\mathrm{BE}} = \frac{1}{\mathrm{e}^{\frac{\hbar \omega}{k_\mathrm{B} T}}-1}.
\end{align}
MD with classical particles samples the first order expansion of \eqref{eq: BE} at high temperature
\begin{align}\label{eq: EQ}
    n_{\mathrm{BE}} = \frac{k_\mathrm{B} T}{\hbar \omega},
\end{align}
i.e. the equipartition law. Since the difference between Eqs.~\eqref{eq: BE} and~\eqref{eq: EQ} is always negative at finite $T$, the thermal conductivity computed with equipartition will be lower than the one computed with Bose-Einstein distribution at the same level of the theory. We have checked how important this difference is for Li\textsubscript{3}ClO, by modifying \texttt{ShengBTE} to allow occupations according to~\eqref{eq: EQ}. The results are shown in Fig.~\ref{fig: BE vs EQ}. As expected being Eq.~\eqref{eq: EQ} a high-temperature expansion of Eq.~\eqref{eq: BE}, the difference is lower at higher temperatures. This contribution alone, that is around than $5\%$ in most of the temperature range, is not sufficient to explain the difference between the BTE-based and MD-based values of $\kappa$.

\section{Data analysis}

\subsection{Estimation of thermal conductivity from MD simulations}

\begin{figure}[tbh]
    \centering
    \includegraphics[width=\columnwidth]{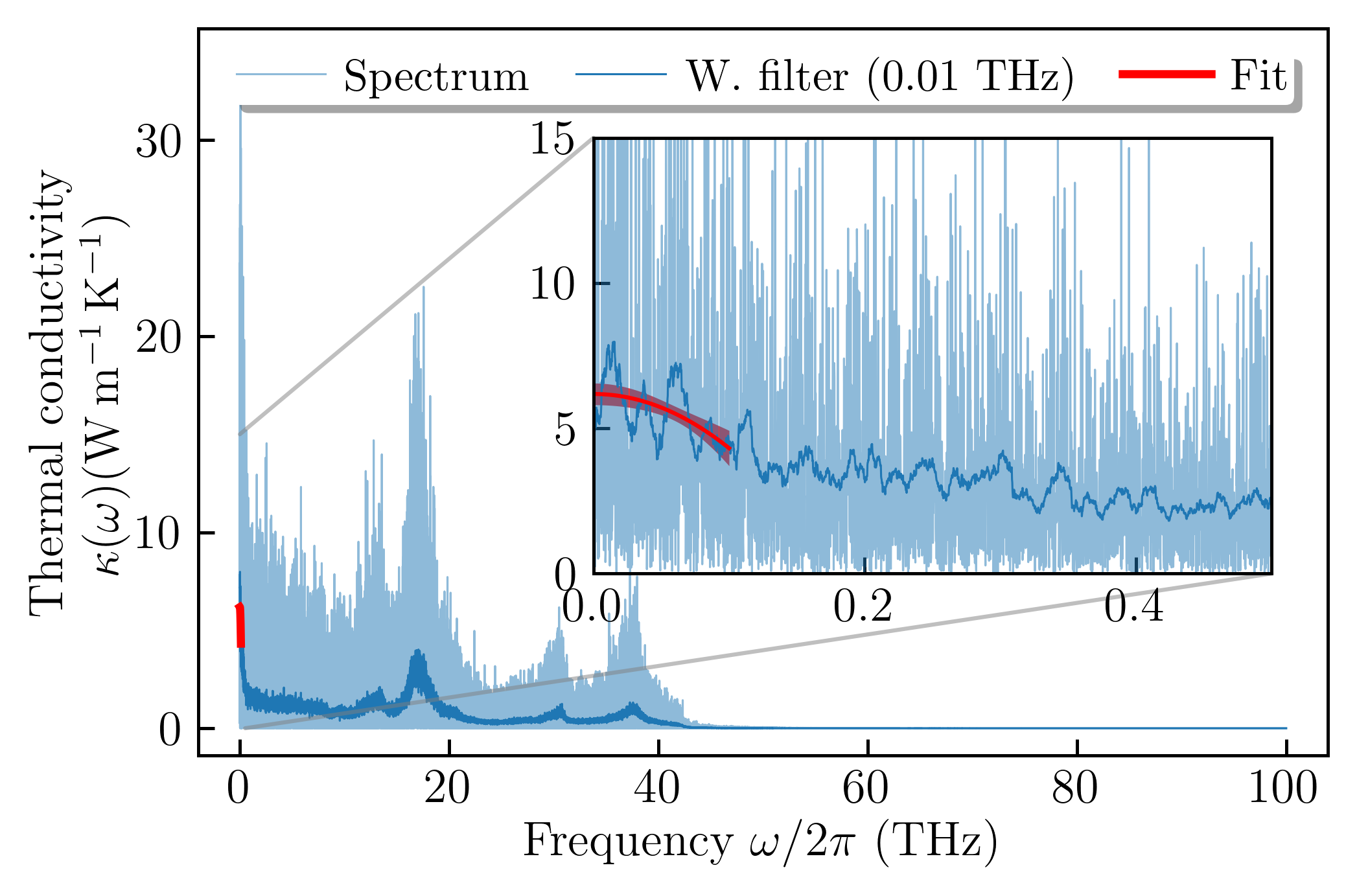}
    \caption{Spectrum of the energy flux decoupled from the mass fluxes for the MD simulation at $T=373\,\mathrm{K}$ and $x=0.005$. The fit to \eqref{eq: S fit} is done on frequencies up to $0.1\,\mathrm{THz}$.}
    \label{fig: spectrum}
\end{figure}

Thermal conductivity is estimated from the power spectrum of the energy flux $S_{EE}$, after removing its coupling to the mass fluxes, $S_{\mathrm{coupl.}}$, i.e.
\begin{align}
    \kappa(\omega) = \frac{\Omega}{6k_\mathrm{B} T^2} \left[
        S_{EE}(\omega) - S_{\mathrm{coupl.}}(\omega) \right].
\end{align}
Details on the method are found in Refs.~\onlinecite{ercole2017accurate} and~\onlinecite{bertossa2019theory}. Being the spectrum an even function of $\omega$, its low frequency part needs to be quadratic. Thus, the first portion (up to $\approx 0.1\,\mathrm{THz}$) of the spectrum is fitted to the quadratic function
\begin{align}\label{eq: S fit}
    \kappa_{\mathrm{fit}}(\omega) & a + b \left(\frac{\omega}{2\pi}\right)^2.
\end{align}
An example ($T=373\,\mathrm{K}$ and $x=0.005$) of the spectrum of the energy flux decoupled from the mass fluxes is shown in Fig.~\ref{fig: spectrum}.

\subsection{Vacancy-dependent GK thermal conductivity}

GK thermal conductivity from FF based MD simulations are fitted for each value of the vacancy concentration $x$. The fitting function is Eq.~\eqref{eq: kappa fitting function}. The values of the fitting parameters $C_{\mathrm{Euck}}$ and $C_{\mathrm{AF}}$, obtained via standard linear regression of $\log(\kappa)$ versus $\log(1/T)$, are shown in Table~\ref{tab: fit}.

\begin{table}[h]
\small
  \caption{Values of the fitting parameters $C_{\mathrm{Euck}}$ (in $\mathrm{W\,m^{-1}}$) and $C_{\mathrm{AF}}$ (in $\mathrm{W\,m^{-1}\,K^{-1}}$) in Eq.~\eqref{eq: kappa fitting function}.}
  \label{tab: fit}
  \begin{tabular}{@{\extracolsep{\fill}}cccccc}
    \toprule
    & $x=0$ & $x=0.005$ & $x=0.02$ & $x=0.05$ & $x=0.1$\\
    $C_{\mathrm{Euck}}$ & $2300 \pm 120$ & $1560 \pm 110$ & $1010 \pm 120$ & $590 \pm 70$ & $320 \pm 90$ \\
    $C_{\mathrm{AF}}$ & $0.2 \pm 0.3$ & $1.5 \pm 0.3$ & $2.1 \pm 0.3$ & $2.1 \pm 0.2$ & $1.9 \pm 0.3$ \\
    \bottomrule
  \end{tabular}
\end{table}

\section{Neural Network model}\label{app: NN}

A Deep Potential--Smooth Edition (DeepPot-SE) Neural Network model~\cite{han2017deep, zhang2018end} is trained on a set of AIMD simulations at different temperatures, enriched with additional configurations chosen by the \texttt{dpgen} software~\cite{dpgen}. The system used for the training of the NN is a ${3\times 3 \times 3}$ supercell of Li\textsubscript{3}ClO with a LiCl pair removed. The Root Mean Squared Errors (RMSEs) of the trained model are $0.11 \pm 0.01\,\mathrm{meV/atom}$ and $15 \pm 3\,\mathrm{meV/\AA}$ for the energy and the forces, respectively. The uncertainty on the RMSEs is the standard deviation of the RMSEs of four different models that differ by the choice of the initial random seed.

\begin{figure}[t]
    \centering
    \includegraphics[width=\columnwidth]{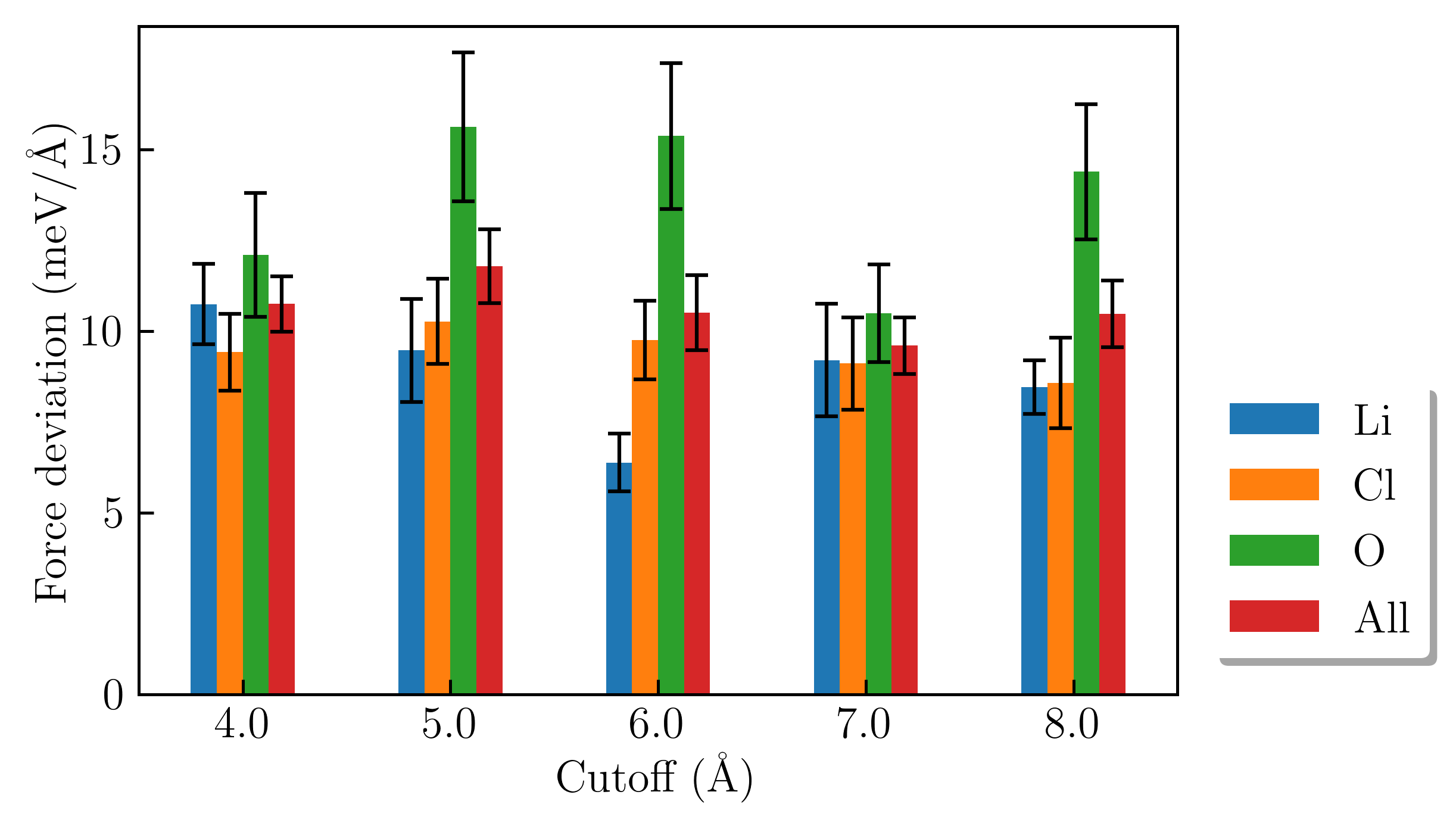}
    \caption{Average force deviations for the locality test. The error bars represent the standard deviation of the force deviation sample.}
    \label{fig: locality-test}
\end{figure}

\begin{figure}[t]
    \centering
    \includegraphics[width=\columnwidth]{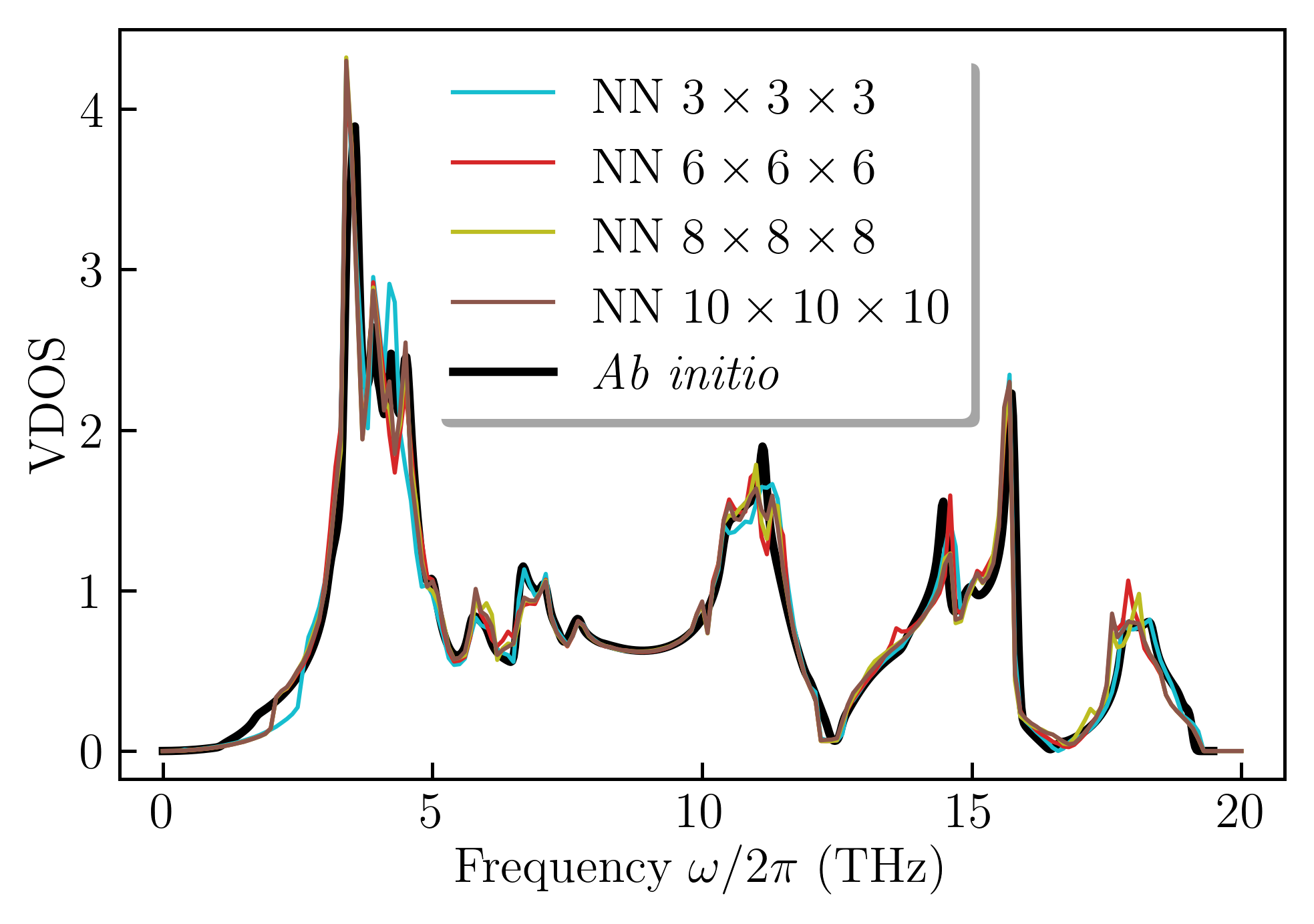}
    \caption{Phonon density of states for different supercell sizes compared with the \emph{ab initio} result.}
    \label{fig: NN phonons}
\end{figure}

The locality test~\cite{bartok2010gaussian} is performed to assess the influence of long-range interactions on the NN model, that are not explicitly considered here. For a given atom, the atoms that are within a chosen radius (in PBCs) from it are kept fixed. A random perturbation is applied to the positions of all the other atoms placed outside the sphere. The force acting on the atom at the centre of the sphere are collected for a number of different random perturbations, computed both with the NN model and \emph{ab initio}. The average deviation between the NN and \emph{ab initio} forces, defined as ${\sigma_f = \sqrt{\langle|\bm{F}_{AI}-\bm{F}_{NN}|^2\rangle}}$, quantifies the dependence of the atom's properties on its neighbours; hence the name of Locality Test. This procedure is carried out for different values of the cutoff and for each atomic species in the system. The results obtained with a ${4 \times 4 \times 4}$ supercell are shown in Fig.~\ref{fig: locality-test}. 
Force deviations are at most compatible with the errors on the training, and exhibit a mildly decreasing dependence on the cutoff of the sphere. The error is computed as the standard error on the average of 10 equivalent configurations. This is consistent with the assumption that the local environment is dominant in determining the properties of an atom, and validates the model we employ.
Phonon density of states is computed and compared to the \emph{ab initio} results for different supercells. The results are satisfactory and are shown in Fig.~\ref{fig: NN phonons}.

\bibliography{main}
\end{document}